\documentclass[pra,twocolumn,superscriptaddress,showpacs,amsmath,amssymb,floatfix]{revtex4}
\usepackage{graphicx,color}
\usepackage{epsfig}
\usepackage{graphicx}
\usepackage{amsmath}
\usepackage{color}
\newcommand{\bra}[1]{\langle{#1} |}
\newcommand{\ket}[1]{|{#1}\rangle  }

\newcommand{\ketbra}[2]{\vert {#1} \rangle \langle{#2}\vert}

\begin{document}

\title{Transmission of classical and quantum information through a\\ 
quantum memory channel with damping}

\author{Antonio D'Arrigo}
\affiliation{CNR-IMM-Uos Catania (Universit\`a), Consiglio Nazionale delle Ricerche, 
Via Santa Sofia 64, 95123 Catania, Italy} 
\affiliation{Dipartimento di Fisica e Astronomia, Universit`a degli Studi Catania, c/o Viale Andrea Doria 6, Ed. 10, 95125 Catania, Italy} 
\author{Giuliano Benenti}
\affiliation{CNISM, CNR-INFM \& Center for Nonlinear and Complex Systems, 
Universit\`a degli Studi dell'Insubria, Via Valleggio 11, 22100 Como, Italy}
\affiliation{Istituto Nazionale di Fisica Nucleare, Sezione di Milano,
via Celoria 16, 20133 Milano, Italy}
\author{Giuseppe Falci}
\affiliation{Dipartimento di Fisica e Astronomia, Universit`a degli Studi Catania, c/o Viale Andrea Doria 6, Ed. 10, 95125 Catania, Italy} 
\affiliation{CNR-IMM-Uos Catania (Universit\`a), Consiglio Nazionale delle Ricerche, 
Via Santa Sofia 64, 95123 Catania, Italy} 
\affiliation{Centro Siciliano di Fisica Nucleare e di Struttura della Materia (CSFNSM), Viale S. Sofia 64, 95123 Catania, Italy}

\begin{abstract}
We consider the transfer of classical and quantum information through a memory
amplitude damping channel. Such a quantum channel is modeled as a damped 
harmonic oscillator, the interaction between the information carriers - a train 
of qubits - and the oscillator being of the Jaynes-Cummings kind. 
We prove that this memory channel is forgetful, so that quantum coding theorems 
hold for its capacities.
We analyze entropic quantities relative to two uses of this channel.
We show that memory effects improve the channel aptitude to transmit both
classical and quantum information, and we investigate the mechanism by which memory
acts in changing the channel transmission properties.
\end{abstract}


\maketitle
\section{Introduction}
Noisy channel coding theorems
developed  in the last fifteen 
years~\cite{shor-95,westmoreland,holevo2,lloyd,devetak} 
are a central topic in 
quantum information theory~\cite{nielsen-chuang,benenti-casati-strini}. 
They provide  quantifiers for the effects of noise in quantum
communication and allow to investigate how it 
is possible to reliably communicate
through quantum channels. These are 
mathematically described in the 
quantum operation ${\cal E}$~\cite{nielsen-chuang,benenti-casati-strini}
formalism: an input state described by
the density matrix $\rho_{\textsf Q}$ yields an output state
$\rho_{\textsf Q}'=\mathcal{E}(\rho_{\textsf Q})$, where the 
quantum channel $\mathcal{E}$ is a linear map. Effects of noise 
are described by a non-unitary $\cal E$.
\\
Such picture can be applied to a variety of situations, for instance
to information storage into a \textit{quantum memory}~\cite{quantum-memory}
or to state transfer from one unit to an other of a quantum 
computer~\cite{nielsen-chuang,benenti-casati-strini}; it may describe 
photons traveling across a fiber~\cite{gisin} in a quantum cryptographic 
system or entanglement distribution between 
different parties~\cite{nielsen-chuang,benenti-casati-strini}.
\\
A central question is: how is it possible to overcome the information 
degradation due to noise, in order to obtain a reliable transmission?
As in the classical realm, in order to protect communication
one may add some redundancy to the information encoding at the channel 
input~\cite{nielsen-chuang,benenti-casati-strini}. 
On the other hand redundancy lowers the channel efficiency in terms 
of information transmitted per channel use. 
So the former question becomes: ``what is the maximum rate of classical 
or quantum information that can be faithfully transmitted?''. 
Classical and quantum capacities~\cite{bennet-shor}, 
defined as the maximum number of bits/qubits that can be reliably 
transmitted per channel use (in the asymptotic limit of an infinite 
number of channel uses), provide the answer to this question.
\\
From the physical side, carriers for quantum or classical 
information are quantum systems. Noise in communication is due to the 
unavoidable interaction of such quantum systems 
with their environment. Since typically a quantum channel has to be used 
many times, this physical perspective makes clear that the 
backaction of the environment may play a crucial role in the determination 
of the maximal rate of faithful information transfer. The 
simplest models for quantum channels are memoryless, that is to say,
at each channel use the environment backaction is negligible 
and the system undertakes the same transformation ${\cal E}_1$.
If we use the channel $N$ times we have that
$\rho_{\textsf Q}^{(N)\prime}={\cal E}_N(\rho_{\textsf Q}^{(N)})$
where ${\cal E}_N={\cal E}^{\otimes N}$. 
On the other hand real systems exhibit correlations among subsequent
uses. For example this can happen when environment correlation times 
are longer than the time between consecutive uses. These kind of channel
are called \textit{memory channel} and for these we have 
${\cal E}_N\neq{\cal E}_1^{\otimes N}$. 
\\
Starting from the seminal work~\cite{palma}, quantum channels with memory 
have attracted increasing attention in the last 
years, see~\cite{hamada,mancini,giovannetti,werner,datta,virmani,njp,rossini,bosonic,spins,memorycavity,gabriela,lupo,banaszek} and references therein. 
Coding theorems have been proved for classes of quantum memory channels~\cite{werner,datta}.
The presence of memory can bring interesting features: 
memory can enhance
the quantum capacity of a channel~\cite{virmani,njp}, and
ensemble of entangled states can improve the memory 
channel aptitude to transmit
classical information~\cite{palma,banaszek}. 
\\
Early models of correlated uses considered 
Markov chains to model memory effects~\cite{palma,hamada,datta}:
on one hand this gave a powerful tool to begin to understand how memory 
effects may change the transmission property of a given channel, on the other
hand often real systems exhibit a dynamics which is not captured by 
simple Markov chains.
For instance solid state systems, which are very promising for quantum 
information purposes thanks to their potential integrability, scalability 
and controllability, are subject to a variety of noise sources producing
low-frequency noise, often exhibiting a typical $1/f$ 
spectrum~\cite{solid-state}. A realistic description of such systems 
may require the Hamiltonian modeling of the sources of noise with memory, 
and in general studying selected Hamiltonian 
models provides basic insight to understand memory effects of 
real system used as quantum channels. 
\\
In this work we consider a Hamiltonian model implementing 
a \textit{memory amplitude damping channel}, and study exactly 
the behavior for two-channel uses.
The amplitude damping channel~\cite{nielsen-chuang,benenti-casati-strini}
models for instance environment-induced relaxation processes from the 
excited state to the ground state of a quantum two-level system. 
It provides an adequate description of systems where relaxation is 
dominant with respect to pure dephasing processes. 
\\
The Hamiltonian model we choose describes qubit (information carriers) 
sequentially interacting with a damped harmonic mode. 
The interaction is of the Jaynes-Cummings type, therefore the whole system 
can be visualized by a qubit-micromaser~\cite{micromaser} system, 
the qubit train being a stream of two-level Rydberg atoms injected at some 
rate into the cavity. 
This model is fundamental in quantum optics, where it 
describes Cavity~\cite{haroche} Quantum Electrodynamics (QED) systems, where 
quantum two level systems are strongly coupled to discrete photon modes in 
high-quality cavities; recently  Circuit~\cite{wallraff} QED systems 
operating in the microwave range have been implemented on a solid-state 
platform and similar architectures have been proposed for the implementation 
of hybrid quantum memories~\cite{quantum-memory}.
A single use yields an an amplitude damping channel~\cite{chen} ${\cal E}_1$, 
provided the cavity is prepared in its ground state.
Memory effects naturally arise when repeated uses are considered 
due the fact that the cavity changes its state after interacting with each
information carrier. The strength of memory depends on the cavity quality 
factor: the greater it is, the stronger are memory effects.
\\
In a previous work we have shown that in the limit of strong dephasing
this environment, although decreasing the information 
transfer per use (the coherent information $I_C$, see Sec. II), 
in a regime where memory effects are present still yields a 
rate of information transfer larger than operating with a low qubit 
injection rate to allow memory to decay.
In this paper we analyze the channel aptitude to transmit
quantum and classical information by considering exactly 
two uses of this channel. We show that the channel is 
\textit{forgetful}~\cite{werner}, 
and therefore usual quantifiers for quantum 
communication satisfy noisy coding theorems~\cite{shor-95,westmoreland,holevo2,lloyd,devetak}. 
A careful study of the above quantities allows to 
conclude that memory effects improve the channel aptitude to transmit both
quantum information, i.e. they may determine a larger $I_C$ per use 
than in the absence of memory. An analogous statement also holds for 
transmission of classical information. 
\\
The paper is organized as follows. In section II we recall the concepts of 
quantum and classical capacities which apply to memoryless channels and
for the class of forgetful memory channels~\cite{werner}.
In section III we present our channel model and demonstrate its forgetfulness.
In section IV we discuss the memoryless limit, whereas in section V we present
numerical results for two channel uses, characterizing the channel behavior 
by suitable entropic quantities; we also investigate the mechanism by which 
memory changes the channel aptitude to transmit classical and quantum 
information. 
Finally in section VI we discuss the significance of our results, 
pointing out the key role of qubit-environment entanglement in 
increasing correlations between successive uses.

\section{Quantum capacities}
We consider a quantum information carrier, which is a quantum system 
$\textsf Q$ whose state is chosen from the ensemble $\{\sigma_1,...,\sigma_K\}$ ($\sigma_i$ belongs 
to the Hilbert space ${\cal B}_1$ of the system),
with a priori probabilities $\{\xi_1,...,\xi_K\}$. 
This quantum source is then 
described by the density operator~\cite{barnum}
\begin{equation}
\rho_{\textsf Q}\,=\,\sum_{i=1}^K\, \xi_i\, \sigma_i.
\label{eq:q-source-density-operator}
\end{equation}
Due to the unavoidable coupling to uncontrollable degrees of
freedom, the transmission of this information is in general noisy.
{Indeed any quantum system interacts
with its environment, thus deviating from the expected evolution.
Therefore
one has to deal with a larger system $\textsf{QE}$,  whose Hamiltonian is 
${\cal H}\,=\,{\cal H}_{\textsf Q}+{\cal H}_{\textsf E}
+{\cal H}_{\textsf QE}$, and describes the system (${\cal H}_{\textsf Q}$), 
the environment (${\cal H}_{\textsf E}$) and their mutual 
interaction (${\cal H}_{\textsf QE}$). 
The reduced evolution of $\textsf Q$ is obtained as a linear, 
completely positive, trace 
preserving (CPT) map~\cite{nielsen-chuang,benenti-casati-strini}
$\mathcal{E}(\rho_{\textsf Q})\,=\,{\rm Tr}_{\textsf{E}}\{\tilde{U}_{\textsf{QE}}\,
 (\rho_{\textsf Q} \otimes \omega_{\textsf E}) \,\tilde{U}^\dag_{\textsf{QE}}\}$. Here
$\omega_{\textsf E}$ is the initial state of the environment, 
$\tilde{U}_{\textsf{QE}}=\exp(-i\tilde{{\cal H}}_{\textsf{QE}} t)$ 
and $\tilde{{\cal H}}_{\textsf{QE}}$ is the
Hamiltonian of $\textsf{QE}$ 
in the interaction picture with respect to ${\cal H}_{\textsf Q}$ and ${\cal H}_{\textsf E}$. The map 
${\cal E}$ is the quantum channel, accounting for the noisy processes
that the information carrier undergoes during transmission.} 
In what follows we consider $N$ channel uses. The quantum source is still
described by (\ref{eq:q-source-density-operator}), where now 
$\sigma_i \in {\cal B}_1^{\otimes N}$, and the $N$ uses map
is denoted by $\mathcal{E}_N(\rho)$.

In this paper we are interested to 
the quantum 
capacity $Q$ and the classical capacity $C$.
The former quantifies the faithful transmission of quantum information, 
measured by the dimension of the largest subspace of the $N$-use input Hilbert
space $\mathcal{B}_1^{\otimes N}$ that can be
reliably transmitted down the channel, in the limit of large $N$.
The classical capacity $C$ gives the maximum amount of classical 
information that can be reliably transmitted per channel use.
That is to say, we wish to identify the largest set of orthogonal 
input states which remain distinguishable (i.e., orthogonal).
In this latter case, the system is not required
to preserve the coherence of superpositions of such input states.
It is clear that $C\ge Q$, since we can always use orthogonal 
states in the subspace reliably transmitted to encode classical
information.

\subsection{Memoryless channels}
We first consider memoryless channels, 
$\mathcal{E}_N=\mathcal{E}_1^{\otimes N}$, that is each single use 
map $\mathcal{E}_1$ acts independently on the system $\textsf Q$.
The quantum capacity $Q$ (measured in qubits 
per channel use) can be
computed as~\cite{lloyd,barnum,devetak}
\begin{equation}
Q \,=\, \lim_{N\to\infty} \frac{Q_N}{N},
\quad \quad
Q_N\,=\,\max_{\rho_{\textsf Q}} I_c(\mathcal{E}_N,\rho_{\textsf Q}),
\label{qinfo}
\end{equation}
\begin{equation}
I_c(\mathcal{E}_N,\rho_{\textsf Q})
\,=\,S[\mathcal{E}_N(\rho_{\textsf Q})]-
S_e(\mathcal{E}_N,\rho_{\textsf Q}).
\label{coinfo}
\end{equation}
Here $S(\rho)=-\mathrm{Tr}[\rho \log_2 \rho]$ is the von
Neumann entropy,
$S_e(\mathcal{E}_N,\rho_{\textsf Q})$ is the \textit{entropy exchange}~\cite{schumacher},
defined as
$$S_e(\mathcal{E}_N, \rho_{\textsf Q}) = S[(\mathcal{I} \otimes \mathcal{E}_N)(|\Psi\rangle\langle \Psi|)],$$
where $|\Psi\rangle$ is any purification of $\rho_{\textsf Q}$.
That is, we consider the system ${\textsf Q}$, described by the density
matrix $\rho_{\textsf Q}$, as a part of a larger quantum system 
${\textsf R}{\textsf Q}$ in a pure state $|\Psi\rangle$; 
the reference system ${\textsf R}$ evolves trivially, according to the identity
superoperator $\mathcal{I}$ whereas 
$\rho_{\textsf Q}=\mathrm{Tr}_{\textsf R} |\Psi\rangle\langle\Psi|$.
The quantity $I_c(\mathcal{E}_N,\rho_{\textsf Q})$ is called 
\textit{coherent information}~\cite{schumachernielsen}
and must be maximized over over all input states $\rho$.

The regularization $N\to\infty$ is necessary since in general
$I_c$ fails to be subadditive~\cite{barnum} and therefore it cannot
be excluded that $Q_N/N>Q_1$. 
The regularization is not necessary if the final state 
of the environment can be reconstructed from the final state of the system.
In this case, referred to as \textit{degradable channels}~\cite{degradable},
the coherent information reduces to a suitable 
conditional entropy, which is subadditive, and
the quantum capacity is given by the 
``single-letter'' formula $Q=Q_1$.
It is worth noticing that for degradable channels the
private classical capacity $C_p$, defined
as the capacity for transmitting classical information 
protected against an eavesdropper~\cite{devetak}, is equal to the quantum
capacity $Q$~\cite{smith}.

The reliability of transmission of a subspace $\mathcal{B}$
through a channel described by the CPT map $\mathcal{E}$
can be measured by the minimum
pure-state fidelity~\cite{footnote}
\begin{equation}
F_p(\mathcal{B},\mathcal{E})\equiv
\min_{|\psi\rangle \in \mathcal{B}}
\langle \psi | \mathcal{E}(|\psi\rangle\langle\psi|)|\psi\rangle.
\end{equation}
Following~\cite{BKN2000} we say that the rate (per channel use) 
$R$ of transmission of subspace dimensions is
achievable with channel $\mathcal{E}_1$ if there exists a sequence
of subspaces $\mathcal{B}^{(N)}$ of $\mathcal{B}_1^{\otimes N}$ such that
\begin{equation}
\limsup_{N\to \infty}
\frac{\log_2 {\rm dim}\left(\mathcal{B}^{(N)}\right)}{N}=R
\end{equation}
and there are coding and decoding CPT maps, $\mathcal{C}^{(N)}$ and
$\mathcal{D}^{(N)}$, such that
\begin{equation}
\lim_{N\to\infty} F_p\left(\mathcal{B}^{(N)},\mathcal{D}^{(N)}
\circ \mathcal{E}_1^{\otimes N}\circ \mathcal{C}^{(N)}\right)=1
\end{equation}
(for the sake of simplicity, we assume that $\mathcal{E}^{\otimes N}$,
$\mathcal{C}^{(N)}$, and $\mathcal{D}^{(N)}$ all act on the same Hilbert
space $\mathcal{B}_1^{\otimes N}$).
The quantum capacity $Q$ of the channel is the supremum of achievable rates $R$.
Therefore, $\forall \varepsilon,\delta>0$,  $\exists N_0\in \mathbb{N}$
such as, $\forall N>N_0$, $\exists \,\mathcal{C}^{(N)},\mathcal{D}^{(N)},
\mathcal{B}^{(N)} \subseteq \mathcal{B}_1^{\otimes N}$,
with $[\log_2 {\rm dim}\left(\mathcal{B}^{(N)}\right)]/{N}\ge Q-\delta$,
and we can transmit each state in the subspace $\mathcal{B}^{(N)}$
with high fidelity, i.e., $F_p\left(\mathcal{B}^{(N)},\mathcal{D}^{(N)}
\circ \mathcal{E}_1^{\otimes N}\circ \mathcal{C}^{(N)}
\right)\ge 1-\varepsilon$.
That is to say, 
$\forall |\psi\rangle\in \mathcal{B}^{(N)}$,
\begin{equation}
\langle \psi | \mathcal{D}^{(N)}\circ \mathcal{E}_1^{\otimes N}
\circ \mathcal{C}^{(N)}
(|\psi\rangle\langle \psi|) |\psi\rangle \ge 1-\varepsilon.
\label{psfcondifion}
\end{equation}

The classical capacity is given by~\cite{holevo2,hausladen,westmoreland}
\begin{equation}
C \,=\, \lim_{N\to\infty} \frac{C_N}{N},
\quad \quad
C_N\,=\,\max_{\{\xi_k,\sigma_k\}} \chi(\{\xi_k,\sigma_k\}),
\label{cinfo}
\end{equation}
\begin{equation}
\chi(\{\xi_k,\sigma_k\})
\,=\,S[\mathcal{E}_N(\rho_{\textsf Q})]-
\sum_k \xi_k S[\mathcal{E}_N(\sigma_k)],
\label{holevo}
\end{equation}
with $\rho_{\textsf Q}$ given by (\ref{eq:q-source-density-operator}) and $\chi$ is the Holevo 
information~\cite{holevo,nielsen-chuang,benenti-casati-strini}.
Therefore, $C_N$ is obtained after maximization of the Holevo 
information over all possible ensembles $\{\xi_k,\sigma_k\}$ which produce $\rho_{\textsf Q}$. 
Note that for the reliable transmission of classical information
the system is not required to preserve relative phases of superpositions 
of different messages.  
Therefore, condition (\ref{psfcondifion}) must be fulfilled not 
for all states in subspace $\mathcal{B}^{(N)}$ but just for
a orthonormal basis of that subspace. Classical capacity is then
obtained as the supremum of achievable rates (per channel use)
of transmission of classical information. 

\subsection{Memory channels}
\label{sec:memorychannel}
When memory effects are taken into account,
${\cal E}_N \neq {\cal E}_1^{\otimes N}$, i.e.,
the channel does not act on each carrier independently.
In this case the regularized coherent and Holevo information
in general only provide upper
bounds on the channel capacities. However, for the class of
\emph{forgetful channels}~\cite{werner}, for which memory effects
decay exponentially with time, a quantum coding theorem showing
that this bounds can be saturated exists~\cite{werner}.

It is useful to express forgetfulness
in an operational way, pointing out the feature
that allows the mapping of a
forgetful channel into a memoryless one, with negligible error.
The key point is the use of a double-blocking strategy.
Following Ref.~\cite{werner} (see also Ref.~\cite{virmani}),
we consider blocks of $N+L$ uses of the channel and
do the actual coding and decoding for the first $N$ uses, ignoring the
remaining $L$ idle uses.
The resulting CPT map $\bar{\mathcal{E}}_{N+L}$
actually acts on density matrices $\rho$ on $\mathcal{B}_1^{\otimes N}$.
Considering $M$ uses of such blocks, 
the corresponding CPT map
$\bar{\mathcal{E}}_{M(N+L)}$ can be approximated, with
arbitrarily small error, by
the memoryless setting $\bar{\mathcal{E}}_{N+L}^{\otimes M}$ if~\cite{werner}:
\begin{equation}
\Vert \bar{\mathcal{E}}_{M(N+L)}(\rho_{\textsf Q}) -
\bar{\mathcal{E}}_{N+L}^{\otimes M}(\rho_{\textsf Q}) 
\Vert_1
\leq h\,(M-1) c^{-L},
\label{eq:forgetful}
\end{equation}
where $\Vert  \rho  \Vert_1\equiv {\rm Tr}\sqrt{\rho^\dagger \rho}$ 
denotes the trace norm~\cite{nielsen-chuang}, $\rho_{\textsf Q}$ is any $MN$ input state, $h$ and $c$ are real parameters,
that  are independent of the input state $\rho_{\textsf Q}$, with $h>0$ and $c>1$. 
Expression (\ref{eq:forgetful})
states that, even though the error committed by replacing
the memory channel itself with the corresponding memoryless channel grows
with the number of blocks $M$, it goes to zero exponentially fast with
the number $L$ of idle uses in a single block.
This key feature of forgetful channels allows the
proof of coding theorems for this class of quantum memory
channels, by mapping them into the corresponding
memoryless channels, for which quantum coding theorems hold~\cite{werner}.
In the following, we will use the wording forgetful channels
for systems satisfying inequality (\ref{eq:forgetful}), independently of the
specific model used in Ref.~\cite{werner}.

Hereafter we will show,
by using arguments similar to the ones used
in Ref.~\cite{virmani} for the classical capacity, 
that the quantum capacity coding theorem holds
for forgetful channels, while we refer 
to~\cite{werner,virmani} for the classical capacity.
The coding theorems seen in the 
previous sections hold for the memoryless channel $\bar{\mathcal{E}}_{N+L}$, 
and therefore a coding-decoding
scheme exists such that $\forall \delta>0$,
$\forall \varepsilon>0$, $\exists M_0\in\mathbb{N}$ such that
$\forall M>M_0$ there exist coding and decoding CPT maps 
$\bar{\mathcal{C}}^{(M)}$ and $\bar{\mathcal{D}}^{(M)}$ 
and subspace $\bar{\mathcal{B}}^{(M)} 
\subseteq \mathcal{B}_1^{\otimes MN}$,
with $[{\log_2 {\rm dim}\left(\bar{\mathcal{B}}^{(M)}\right)}]/({MN})
\ge Q-\delta$,
and, $\forall |\psi\rangle\in \bar{\mathcal{B}}^{(M)}$,
\begin{equation}
\langle \psi | \bar{\mathcal{D}}^{(M)}\circ 
\bar{\mathcal{E}}_{N+L}^{\otimes M}
\circ \bar{\mathcal{C}}^{(M)}
(|\psi\rangle\langle \psi|) |\psi\rangle \ge 1-\varepsilon.
\label{psfcondition2}
\end{equation}

We will prove that the above coding-decoding scheme works well 
also for a forgetful memory channel. 
For this purpose, it is useful to observe 
that (\ref{psfcondition2}) is fulfilled provided that
\begin{equation}
\Vert |\psi\rangle\langle \psi|
-\bar{\mathcal{D}}^{(M)}\circ 
\bar{\mathcal{E}}_{N+L}^{\otimes M}
\circ \bar{\mathcal{C}}^{(M)}
(|\psi\rangle\langle \psi|)  
\Vert_1 \le \frac{\varepsilon}{2}
\label{psfcondition4}
\end{equation}
$\forall |\psi\rangle\in \bar{\mathcal{B}}^{(M)}$~\cite{notenorm}.
Note that the first member of the inequality (\ref{psfcondition4})
is just a measurement of the error probability in transmitting the generic state $|\psi\rangle$. 
To prove the validity of the coding-decoding scheme for the 
memory channel, we will show that this error is small also for
the channel $\bar{\mathcal{E}}_{M(N+L)}$, namely:
\begin{equation}
\Vert |\psi\rangle\langle \psi|-
\bar{\mathcal{D}}^{(M)}\circ 
\bar{\mathcal{E}}_{M(N+L)}
\circ \bar{\mathcal{C}}^{(M)}
(|\psi\rangle\langle \psi|)  
\Vert_1
\label{psfcondition3}
\end{equation}
can be made arbitrarily small. 
Due to the triangle inequality, an upper bound to the quantity 
in (\ref{psfcondition3}) is provided by
\begin{eqnarray}
&&\Vert |\psi\rangle\langle \psi|
-\bar{\mathcal{D}}^{(M)}\circ
\bar{\mathcal{E}}_{N+L}^{\otimes M}
\circ \bar{\mathcal{C}}^{(M)}
(|\psi\rangle\langle \psi|)             
\Vert_1\,
\nonumber\\
&&+\Vert \bar{\mathcal{D}}^{(M)}\circ 
\bar{\mathcal{E}}_{N+L}^{\otimes M}
\circ \bar{\mathcal{C}}^{(M)}
(|\psi\rangle\langle \psi|)\, 
\nonumber \\
&&\hspace{0.8cm}-\bar{\mathcal{D}}^{(M)}\circ 
\bar{\mathcal{E}}_{M(N+L)}
\circ \bar{\mathcal{C}}^{(M)}
(|\psi\rangle\langle \psi|) 
\Vert_1.
\label{eq:triangle}
\end{eqnarray}
Knowing that (\ref{psfcondition4}) holds, we have to prove 
 that the second term in (\ref{eq:triangle})
can be made arbitrarily  
small $\forall |\psi\rangle\in \bar{\mathcal{B}}^{(M)}$.
Therefore it is sufficient to prove that 
\begin{equation}
\Vert 
\bar{\mathcal{D}}^{(M)}\circ 
\bar{\mathcal{E}}_{N+L}^{\otimes M}(\rho_{\textsf Q})
-\bar{\mathcal{D}}^{(M)}\circ
\bar{\mathcal{E}}_{M(N+L)}(\rho_{\textsf Q})
\Vert_1
\label{eq:dbs-error}
\end{equation}
can be made small for any input state $\rho_{\textsf Q}\, \in \, \mathcal{B}_1^{\otimes MN}$.
Due to the contractivity of trace-preserving 
quantum operations~\cite{nielsen-chuang},
this latter quantity is upper bounded by  
\begin{equation}
\Vert 
\bar{\mathcal{E}}_{N+L}^{\otimes M}(\rho_{\textsf Q})
-\bar{\mathcal{E}}_{M(N+L)}(\rho_{\textsf Q})
\Vert_1
\label{psfcondition5}
\end{equation}
which in its turn is upper bounded by $h(M-1)c^{-L}$ for a forgetful 
channel (\ref{eq:forgetful}).
Then, quantity (\ref{psfcondition3}) is upper bounded by the sum of the left part of
(\ref{psfcondition4}) and by $h(M-1)c^{-L}$:
\begin{eqnarray}
&&\hspace{0cm}\Vert |\psi\rangle\langle \psi|-
\bar{\mathcal{D}}^{(M)}\circ 
\bar{\mathcal{E}}_{M(N+L)}
\circ \bar{\mathcal{C}}^{(M)}
(|\psi\rangle\langle \psi|)  
\Vert_1 \leq \nonumber\\
&&\hspace{0.5 cm} \Vert |\psi\rangle\langle \psi|-
\bar{\mathcal{D}}^{(M)}\circ 
\bar{\mathcal{E}}^{\otimes M}_{(N+L)}
\circ \bar{\mathcal{C}}^{(M)}
(|\psi\rangle\langle \psi|)\Vert_1 \, +\nonumber\\
&&\hspace{1 cm} h(M-1)c^{-L}.
\label{psfconditionlast}
\end{eqnarray}
As it is stressed in Ref.~\cite{virmani}, it is not a priori clear how the 
right part of inequality (\ref{psfconditionlast}) can be made arbitrarily small.
Indeed to make small the first term in the right part of (\ref{psfconditionlast}),
that is the error of the product code, one needs a large number of block
uses $M$, so that (\ref{psfcondition4}) holds; but a large 
M pushes up the the second term in the right part of (\ref{psfconditionlast}),
since this term is proportional to $M$. Nevertheless, thanks to exponential
dependence on the idle uses $L$, both terms can be simultaneously
made arbitrarily small, for $M$ large enough. For example,
if $L=\varepsilon M$ one simply find that there exists a given $M_0'$ such that
$\forall M > M_0'$ quantity (\ref{psfcondition3}) is less than $\varepsilon$~\cite{noteM0}.

{It is worth noticing that we demonstrate that 
the \textit{memory} channel $\bar{\mathcal{E}}_{N+L}$, 
can faithful transmit any amount of quantum information supported
by the corresponding \textit{memoryless} channel, 
whose quantum capacity is given by applying eq. (\ref{qinfo}) to the map ${\cal E}_N$.
But one can think that the maximum faithful rate achievable by $\mathcal{E}_{N+L}$ is greater
than the one of 
$\bar{\mathcal{E}}_{N+L}$, because with a double-block strategy
one throws away the information related to the idle uses. }
But, it is simple to demonstrate 
that (see Appendix~\ref{app:doubleBlockStrategy}):
\begin{eqnarray}
&& I_c(\rho^{(N+L)},{\cal E}_{N+L}) \le \nonumber \\
&&\hspace{1cm} I_c(\rho^{(N+L)},\bar{\mathcal{E}}_{N+L}) \, + \, L\log_2 \dim({\cal H}_1).
\end{eqnarray}
Therefore, the quantum capacity (Eq. (\ref{qinfo})) of the two channels
${\cal E}_{N+L}$ and $\bar{\mathcal{E}}_{N+L}$ coincides when
$\lim_{N\to\infty} (L/N) = 0$. To fulfill this condition, 
taking into account that we have set $L=\varepsilon M$, we can 
choose $N=M^\alpha$, with $\alpha>1$.

\section{The model}
\label{sec:model}
We consider a stream of $N$ qubits (the \emph{system}, $\textsf Q$) 
interacting with a \textit{structured environment}, composed by a harmonic oscillator 
(the \emph{local ``unconventional'' environment}~\cite{unconventional}, $\textsf O$),
which in turn is damped due to the coupling with a  
\emph{reservoir}. The overall Hamiltonian reads 
\begin{eqnarray}
&&{\cal H}(t)={\cal H}_{0}+{\cal H}_{\textsf{QO}}+\delta {\cal H}, \;
                 \nonumber \\
&&{\cal H}_0={\cal H}_{\textsf Q}+ {\cal H}_{\textsf O}= 
                          -\frac{\omega}{2}\sum_{k=1}^N \sigma_z^{(k)}\,+\,
                          \nu\left(a^\dagger a+\frac{1}{2}\right),
                 \nonumber \\
&&{\cal H}_{\textsf{QO}}=\lambda \sum_{k=1}^N f_k(t) \left( a^\dagger \sigma_-^{(k)}+
a \sigma_+^{(k)}\right).
\label{model}
\end{eqnarray}
Here $\sigma_z=\ketbra{g}{g}-\ketbra{e}{e}$, being $\ket{g}$ and $\ket{e}$ the two
qubit eigenstates, and $\sigma_+=\ketbra{e}{g}$, $\sigma_-=\ketbra{g}{e}$ are the 
qubit rising and lowering operators; $a^\dag$ and $a$ are the creation and annihilation
operators for the harmonic oscillator.
The qubits-oscillator interaction ${\cal H}_{\textsf{QO}}$ is of the
Jaynes-Cummings kind, and 
we take $\lambda$ real and positive and set $\hbar=1$. 
The coupling is switchable:
$f_k(t)=1$ when qubit $k$ is inside the channel (transit time 
$\tau_p$), $f_k(t)=0$ otherwise. 
The term $\delta {\cal H}$ describes both the reservoir's Hamiltonian
and the local environment-reservoir interaction 
and causes damping of the oscillator, within a   
dissipation time scale $\tau_d$. 
Two consecutive qubits entering the channel are separated 
by the time interval $\tau$.

We focus on the resonant regime $\nu \sim \omega$,
arguably the most significant when describing the coupling 
to modes inducing damping. We work in the
interaction picture, where 
the effective Hamiltonian is given by
$\tilde{{\cal H}}=
e^{i{\cal H}_0t} ({\cal H}_{\textsf{QO}} + \delta {\cal H}) e^{-i{\cal H}_0 t}$
(we will 
omit the tilde from now on).

We study the regime in which the transit time $\tau_p$ is smaller 
than the separation time $\tau$ between two consecutive qubits entering 
the channel. Such low-rate regime allows us to neglect 
collective effects such as superradiance, which are expected
to be detrimental for the transmission of classical or
quantum information.

We assume that oscillator 
damping is described by 
the standard master equation (obtained after
tracing over the reservoir) 
\begin{equation}
\dot{\rho}_{\textsf O}\,=\,
\Gamma \, \left(a\, \rho_{\textsf O}\, a^\dagger -
\frac{1}{2} a^\dagger a \,\rho_{\textsf O} -
\frac{1}{2} \rho_{\textsf O}\, a^\dagger a \right).
\label{eq:masterdamping}
\end{equation}
The asymptotic decay (channel reset) to the ground state $|0\rangle$ 
takes place with rate $\Gamma$, so that $\tau_d=1/\Gamma$.
We introduce the memory parameter 
$\mu\equiv \tau_d/(\tau+\tau_d)$.
By definition $0<\mu<1$: fast decay $\tau_d\ll \tau$
yields the 
memoryless limit $\mu\ll 1$, whereas $\mu\lesssim 1$
when memory effects come into play.

We studied model (\ref{model})--(\ref{eq:masterdamping}) in 
Ref.~\cite{memorycavity}, in the case in which an additional 
mechanism of strong dephasing (without relaxation) 
not included in (\ref{eq:masterdamping}) is added.
It was shown that, while the coherent information
per channel use is a decreasing function of the degree of memory, 
the quantum transmission rate, defined as the number of qubits 
that can be reliably transmitted per unit of time, is enhanced
by memory effects. Hereafter we focus on the original model
(\ref{model})--(\ref{eq:masterdamping}), which is computationally
much more demanding but also more interesting, since the interaction
with the local environment can entangle initially separable input
qubits.

\subsection{Forgetfulness}
\label{sect:forgetfulness}
As discussed in Sec.~(\ref{sec:memorychannel}), to prove
that quantum and classical capacities are given by Eqs.~(\ref{qinfo})
and (\ref{cinfo}) it is sufficient to prove 
that (\ref{eq:forgetful}) holds.
Due to the triangle inequality of trace norm, it is sufficient to 
show the validity of inequality (\ref{eq:forgetful}) for $M=2$ (then it follows 
that this inequality holds for any $M>2$, see Appendix \ref{app:forgetfulness}):
\begin{equation}
\Vert \bar{\mathcal{E}}_{2(N+L)}(\rho_{\textsf Q}) -
\bar{\mathcal{E}}_{N+L}^{\otimes 2}(\rho_{\textsf Q}) \Vert_1
\leq h c^{-L}.
\label{eq:forgetful2} 
\end{equation}

Thanks to contractivity of trace-preserving 
quantum operations~\cite{nielsen-chuang},  
we can bound from above the left-hand side of 
(\ref{eq:forgetful2}) with 
\begin{equation}
\Vert \bar{\mathcal{F}}_{2(N+L)}(\rho_{\textsf Q}\otimes \rho_{\textsf O}) -
\bar{\mathcal{F}}_{N+L}^{\otimes 2}(\rho_{\textsf Q}\otimes 
\rho_{\textsf O}) \Vert_1,
\label{eq:forgetful3} 
\end{equation}
where 
$\rho_{\textsf O}=|0\rangle\langle 0|$ is the initial 
(ground) state of the oscillator and  
quantum operations $\bar{\mathcal{F}}_{j(N+L)}$ are such that, 
for any integer $j$, 
\begin{equation}
\bar{\mathcal{E}}_{j(N+L)}(\rho_{\textsf Q})=
{\rm Tr}_{\textsf O}[\bar{\mathcal{F}}_{j(N+L)}
(\rho_{\textsf Q}\otimes \rho_{\textsf O})].
\end{equation}
Here ${\rm Tr}_{\textsf O}$ denotes partial trace over the oscillator 
(we remind the reader that the partial trace is a trace-preserving
quantum operation).
Note that $\bar{\mathcal{F}}$ is non-unitary since the 
oscillator is damped.
The quantity in Eq.~(\ref{eq:forgetful3}) can also be written as
\begin{equation}
\Vert \bar{\mathcal{F}}_{N+L}(\rho_{\textsf{QO}}^\prime) -
\bar{\mathcal{F}}_{N+L}[\pi_0 (\rho_{\textsf{QO}}^\prime)]\Vert_1,
\label{eq:forgetful4}
\end{equation}
where $\rho_{QO}^\prime$ denotes the, generally entangled, 
state of system ($2N$ qubits) and oscillator, after the first $N+L$
channel uses ($L$ idle uses follow the crossing of the channel 
by the first $N$ qubits). The operator 
\begin{equation}
\pi_0(\rho_{\textsf{QO}}')={\rm Tr}_{\textsf{O}}[\rho_{\textsf{QO}}']\otimes\ketbra{0}{0}
\label{eq:memoryless-channel-resetting}
\end{equation}  
removes any correlation established between the qubits and the oscillator, 
and resets the oscillator to its ground state (memoryless setting). In the following we call
$\rho_{\textsf{Q}}'={\rm Tr}_{\textsf{O}}[\rho_{\textsf{QO}}']$
the state of the $2N$-qubit system after the ressetting operation $\pi_0$.

We take again advantage of the fact that trace preserving operations
are contractive to upper bound the quantity in (\ref{eq:forgetful4}) with
\begin{equation}
\Vert \rho_{\textsf{QO}}^\prime -
\pi_0 (\rho_{\textsf{QO}}^\prime)\Vert_1=\Vert \rho_{\textsf{QO}}^\prime -
\rho_{\textsf{Q}}'\otimes\ketbra{0}{0}\Vert_1.
\label{eq:dis-1}
\end{equation}
We also define the state:
\begin{equation}
  \tilde{\rho}_{\textsf{Q}}^\prime= \frac{1}{w_0'}\bra{0}\rho_{\textsf{QO}}^\prime\ket{0},
\end{equation}
where $w_n'={\rm Tr}_{\textsf{QO}}[\Pi_n\, \rho_{\textsf{QO}}'\,\Pi_n]$ and
$\Pi_n\equiv \ketbra{n}{n}$:
$w_n'$ denotes the population, for the state $\rho_{\textsf{QO}}'$, 
of the $n$-th energy eigenstate of the harmonic oscillator.
By using the triangle inequality we can write:
\begin{eqnarray}
&&\hspace{0cm}\Vert \rho_{\textsf{QO}}^\prime -
\rho_{\textsf{Q}}'\otimes\ketbra{0}{0}\Vert_1\le\nonumber\\
&&\hspace{0.5cm}\Vert \rho_{\textsf{QO}}^\prime -
\tilde{\rho}_{\textsf{Q}}'\otimes\ketbra{0}{0}\Vert_1+\Vert \tilde{\rho}_{\textsf{Q}}^\prime -
\rho_{\textsf{Q}}'\Vert_1.
\label{eq:dis-2}
\end{eqnarray}
We start by considering the first term in the right member of (\ref{eq:dis-2}).
Using Uhlmann's fidelity,
$F(\rho,\rho^\prime)={\rm Tr}\sqrt{\rho^{1/2} \rho^\prime \rho^{1/2}}$,
and the inequality
$\Vert \rho-\rho'\Vert_1\leq 
2\sqrt{1-[F(\rho,\rho^\prime)]^2}$~\cite{nielsen-chuang} 
between this fidelity and the trace 
norm, we obtain
\begin{equation}
\begin{array}{c}
\Vert \rho_{\textsf{QO}}^\prime -
\tilde{\rho}_{\textsf Q}^\prime\otimes |0\rangle\langle 0|\Vert_1
\leq 2\sqrt{1-\{F[\rho_{\textsf{QO}}^\prime,\tilde{\rho}_{\textsf Q}^\prime \otimes
|0\rangle\langle 0|]\}^2}.
\end{array}
\label{eq:populationsdiff-1}
\end{equation}
Now we observe that:
\begin{eqnarray}
&&(\tilde{\rho}_{\textsf Q}^\prime \otimes|0\rangle\langle 0|)^{1/2} \cdot
  \rho_{\textsf{QO}}^\prime  \cdot 
  (\tilde{\rho}_{\textsf Q}^\prime \otimes|0\rangle\langle 0|)^{1/2}\,=\nonumber\\
&&\hspace{0.5cm} (\tilde{\rho}_{\textsf Q}^\prime)^{1/2} \cdot (I_{\textsf Q}\otimes|0\rangle\langle 0|) 
  \rho_{\textsf{QO}}^\prime (I_{\textsf Q}\otimes|0\rangle\langle 0|)
 \cdot (\tilde{\rho}_{\textsf Q}^\prime)^{1/2}\,=\nonumber\\
&&\hspace{0.5cm} (\tilde{\rho}_{\textsf Q}^\prime)^{1/2} \cdot w_0'\, 
  \tilde{\rho}_{\textsf{Q}}^\prime \otimes|0\rangle\langle 0|
 \cdot (\tilde{\rho}_{\textsf Q}^\prime)^{1/2}\,=\nonumber\\
&&\hspace{0.5cm}  w_0'\, (\tilde{\rho}_{\textsf Q}^\prime)^{2} \otimes|0\rangle\langle 0|.
 \end{eqnarray}
By substituting this equality into Eq.~(\ref{eq:populationsdiff-1}), we have
\begin{equation}
\begin{array}{c}
 \hspace{-2cm}\Vert \rho_{\textsf{QO}}^\prime -
 \tilde{\rho}_{\textsf Q}^\prime\otimes |0\rangle\langle 0|\Vert_1
 \leq 2\sqrt{1-w_0^\prime}.
 \end{array}
\label{eq:dis-3}
\end{equation}
We now consider the second term of the right member of Eq.~(\ref{eq:dis-2}).
We observe that we can write $\rho_{\textsf Q}^\prime$ as
\begin{equation}
\rho_{\textsf{Q}}^\prime = \sum_{n=0}^\infty \bra{n}\rho_{\textsf{Q}}^\prime\ket{n} =
\sum_n w_n^\prime\rho_{\textsf{Q}_n}^\prime,
\end{equation}
where 
$\rho_{\textsf{Q}_n}^\prime=\frac{1}{w_n^\prime}\bra{n}\rho_{\textsf{Q}}^\prime\ket{n}$;
consequently, $\tilde{\rho}_{\textsf{Q}}^\prime=\rho_{\textsf{Q}_0}^\prime$.
Therefore~\cite{nielsen-chuang}
\begin{eqnarray}
&&\hspace{-0.5cm}\Vert \tilde{\rho}_{\textsf{Q}}^\prime -\rho_{\textsf{Q}}'\Vert_1\le
\sum_{n=0}^\infty w_n^\prime \Vert \rho_{\textsf{Q}_n}^\prime -\rho_{\textsf{Q}_0}^\prime\Vert_1=\nonumber\\
&&\hspace{-0.2cm}
\sum_{n\ge 1}^\infty w_n^\prime \Vert \rho_{\textsf{Q}_n}^\prime -\rho_{\textsf{Q}_0}^\prime\Vert_1
\le 2 \sum_{n\ge 1}^\infty w_n^\prime=2(1-w_0^\prime)\le\nonumber\\
&&\hspace{-0.0cm} 2\sqrt{1-w_0^\prime}
\label{eq:dis-4}
\end{eqnarray}
Adding inequalities (\ref{eq:dis-3}) and (\ref{eq:dis-4}),
we obtain
\begin{eqnarray}
&&\hspace{0cm}\Vert \rho_{\textsf{QO}}^\prime -
\rho_{\textsf{Q}}'\otimes\ketbra{0}{0}\Vert_1\le 4\sqrt{1-w_0^\prime}.
\label{eq:dis-5}
\end{eqnarray}
Moreover we observe that
\begin{equation}
\begin{array}{c}
\hspace{-0.7cm}\sqrt{1-w_0^\prime}=\sqrt{\sum_{n=1}^\infty w_n^\prime}
\le \sqrt{\sum_{n=0}^\infty n w_n^\prime}
\\
\hspace{0.7cm}=\sqrt{\langle a^\dagger a \rangle [(N+L)\tau]}
=e^{-L\Gamma\tau/2}
\sqrt{\langle a^\dagger a \rangle (N\tau)}.
\end{array}
\label{eq:populationsdiff}
\end{equation}

Since $\langle a^\dagger a \rangle$ can at most grow by one
as a result of the unitary interaction with a qubit while 
it drops exponentially at rate $\Gamma$ due to dissipation, we have
\begin{equation}
\langle a^\dagger a \rangle [(k+1)\tau] \le
\langle a^\dagger a \rangle (k\tau)e^{-\Gamma\tau}+1.
\end{equation}
Therefore,
$\langle a^\dagger a \rangle (k\tau) < B$ for any $k$, with
$B=1/(1-e^{-\Gamma\tau})$.
This inequality can be inserted into Eq.~(\ref{eq:dis-5})-(\ref{eq:populationsdiff}),
thus obtaining
\begin{equation}
\Vert \rho_{\textsf{QO}}^\prime -\rho_{\textsf Q}'\otimes 
|0\rangle\langle 0| \Vert_1
< 4 \sqrt{B} e^{-L\Gamma\tau/2},
\end{equation}
which implies that inequality (\ref{eq:forgetful2}) is fulfilled
and therefore quantum and classical capacities can be computed 
from the maximization (\ref{qinfo}) and (\ref{cinfo}) of 
coherent and Holevo information, respectively.

Note that our proof is valid independently of the relation 
between the relevant time scales involved in model (\ref{model}),
namely $\tau_p$, $\tau$, and $\tau_d$. 

\section{The memoryless limit}
It is instructive to consider the memoryless limit
$\tau_d\ll \tau$ ($\mu\ll 1$):
damping acts as 
a built-in reset for the oscillator to its ground state 
$\rho_{\textsf O}(0)=|0\rangle\langle 0|$ after each channel use.
In what follows, we show that 
for $\tau_d\ll \tau$ our model reduces to a memoryless
amplitude-damping channel. 

\subsection{Without damping}
If we further assume that $\tau_p\ll \tau_d$, we can 
neglect nonunitary effects in the evolution of the system and 
the oscillator during the transit time $\tau_p$. 

We consider a generic single-qubit input state,
\begin{equation}
\rho_{\textsf Q_1}(0)=
(1-p) |g\rangle\langle g|+
r |g\rangle \langle e| +
r^\star |e\rangle \langle g| +
p |e\rangle \langle e|,
\label{eq:1qubit}
\end{equation}
with 
$p$ real and $|r|\le \sqrt{p(1-p)}$.
Given the initial, separable qubit-oscillator state
$\rho_{\textsf Q_1}(0)\otimes \rho_{\textsf O}(0)$, we have
\begin{equation}
\mathcal{E}_1[\rho_{\textsf Q_1}{(0)}]=
{\rm Tr}_{\textsf O}
\{U(\tau_p)[\rho_{\textsf Q_1}(0)\otimes \rho_{\textsf O}(0)]U^\dagger(\tau_p)\},
\label{eq:rho1memoryless}
\end{equation}
with $U(\tau_p)$ unitary time-evolution operator determined by the
undamped Jaynes-Cummings Hamiltonian:
\begin{equation}
\left\{                                           \begin{array}{l}
U(\tau_p) |g, 0\rangle = |g,0\rangle,
\\
U(\tau_p) |e, 0\rangle = \cos(\lambda \tau_p) |e,0\rangle
-i \sin(\lambda \tau_p) |g,1\rangle,
\end{array}
\right.
\end{equation}
with $\{|0\rangle, |1\rangle,...\}$ eigenstates of the harmonic oscillator
and $\lambda$ frequency of the Rabi oscillations between levels 
$|e,0\rangle$ and $|g,1\rangle$.
It is then easy to obtain~\cite{chen},
in the $\{|g\rangle,|e\rangle\}$ basis,
\begin{equation}
\mathcal{E}_1[\rho_{\textsf Q_1}{(0)}]=
\left[
\begin{array}{cc}
 1-p\cos^2(\lambda\tau_p) & r\cos(\lambda\tau_p)\\
 r^\star\cos(\lambda\tau_p) & p\cos^2(\lambda \tau_p)
\end{array}
\right].
\label{eq:rho1memoryless2}
\end{equation}
Eq.~(\ref{eq:rho1memoryless2}) corresponds to
an \emph{amplitude-damping channel}:
\begin{eqnarray}
&&\hspace{-0.7cm}\mathcal{E}_1[\rho_{\textsf Q_1}(0)]= 
\sum_{k=0}^1 E_k \rho_{\textsf Q_1}(0) E_k^\dagger,
\nonumber\\
&&\hspace{-0.7cm} E_0=|g\rangle \langle g|+
\sqrt{\eta}\, |e\rangle\langle e|,
\quad
E_1=\sqrt{1-\eta}\, |g\rangle\langle e|,
\label{eq:amplitudedamping}
\end{eqnarray}
with $E_0$, $E_1$ Kraus operators~\cite{nielsen-chuang,benenti-casati-strini}
and $\eta=\cos^2(\lambda \tau_p)\in [0,1]$. 

This channel is degradable~\cite{giovannettifazio}
and therefore to compute its quantum capacity it is sufficient
to maximize the coherent information over a single use of the channel. 
Maximization is achieved 
by diagonal states ($r=0$) and one obtains~\cite{giovannettifazio} 
\begin{equation}
\label{QmemorylessCapacity}
Q=\left\{
\begin{array}{ll}
\max_{p\in [0,1]}
\{H_2(\eta p)-H_2[(1-\eta)p]\} & {\rm if}\; \eta\ge \frac{1}{2},\\\\
0 & {\rm if}\; \eta\le \frac{1}{2},
\end{array}
\right.
\end{equation}
where $H_2(x)=-x\log_2 x -(1-p) \log_2 (1-x)$ is the binary
Shannon entropy.
Since the memoryless amplitude-damping channel is degradable, 
the private classical capacity $C_p$ is equal to the quantum 
capacity $Q$~\cite{smith}.

The classical capacity $C_1$, that is, the maximum amount of 
classical information that can be reliably transmitted using only
encodings that are not entangled over successive uses of the 
channel, is given by~\cite{giovannettifazio}
\begin{equation}
C_1=
\max_{p\in [0,1]}
\left\{H_2(\eta p)-H_2\left(
\frac{1+\sqrt{1-4\eta(1-\eta)p^2}}{2}
\right)\right\}.
\label{eq:C1memoryless}
\end{equation} 
The quantity $C_1$ is a lower bound for the classical capacity $C$.

\subsection{With damping}
In this section we show that our memoryless model always 
reduces to an amplitude-damping channel, independently of
the relation between the time scales $\tau_p$ and $\tau_d$.

The non-unitary evolution of qubit and damped oscillator up
to the crossing time time $t=\tau_p$ is described by the 
Lindblad master equation~\cite{lindblad}
\begin{eqnarray}
&&\hspace{-1cm} \frac{d \rho_{\textsf{QO}}}{d t}= 
-i [{\cal H}_1,\rho_{\textsf{QO}}]\,+\nonumber\\
&&\hspace{1cm}L \rho_{\textsf{QO}} L^\dagger - 
\frac{1}{2} L^\dagger L \rho_{\textsf{QO}} -
\frac{1}{2} \rho_{\textsf{QO}} L^\dagger L,
\label{eq:Lindblad1q}
\end{eqnarray}
where ${\cal H}_1$ is the interaction picture
Hamiltonian (\ref{model}) for a single channel use, 
\begin{equation}
{\cal H}_1=\lambda ( a^\dagger \sigma_-+
a \sigma_+),
\end{equation}
and the Lindblad operator 
\begin{equation}
L=\sqrt{\Gamma} (I_{\textsf s} \otimes a)
\end{equation}
acts trivially on the qubit while damping the oscillator
according to Eq.~(\ref{eq:masterdamping}).

The master equation (\ref{eq:Lindblad1q}) can be solved 
analytically~\cite{chen}, see Appendix~\ref{app:master1q}.
We end up with the amplitude-damping channel (\ref{eq:amplitudedamping}),
but with a $\Gamma$-dependent parameter 
\begin{equation}
\eta(\Gamma)=e^{-\Gamma\tau_p/2}
\left[\frac{\Gamma}{z}\sinh\left(\frac{z\tau_p}{4}\right)
+\cosh\left(\frac{z\tau_p}{4}\right)\right]^2,
\label{eq:etagamma}
\end{equation}
where $z\equiv \sqrt{\Gamma^2-16\lambda^2}$.
Note that for $\Gamma=0$ we recover the undamped result,
$\eta(0)=\cos^2(\lambda \tau_p)$.

For weak damping, we expand Eq.~(\ref{eq:etagamma})
up to first order in $\Gamma/\lambda$, thus obtaining
\begin{equation}
\eta(\Gamma)=\eta(0)+\frac{\Gamma}{4\lambda}
[\sin(2\lambda \tau_p)-
2\lambda\tau_p\cos^2(\lambda\tau_p)].
\end{equation}
Depending on the value of $\lambda\tau_p$ 
we can have either $\eta(\Gamma)>\eta(0)$ or $\eta(\Gamma)<\eta(0)$.
In particular, in the case of nearly ideal transmission
$\lambda\tau_p\ll 1$ ($\eta$ close to $1$), we have, to first
order in $\lambda\tau_p$,
\begin{equation}
\eta(\Gamma)-\eta(0)=\frac{\Gamma}{6\lambda}(\lambda\tau_p)^3>0.
\end{equation}
We can in this case conclude that a small amount of damping improves
the channel capacities $Q$ and $C_1$
(indeed $Q$ and $C_1$ are growing functions of $\eta$~\cite{giovannettifazio}).
This result is to some extent counterintuitive, as damping can 
enhance the quality of the channel.  

It is interesting to examine also the overdamped limit 
$\Gamma/\lambda\to\infty$.
In this case we obtain from 
Eq.~(\ref{eq:etagamma}) that 
$\eta(\Gamma\to\infty)\to 1$, so that the channel becomes
noiseless (capacities $Q,C\to 1$). 
This result is a consequence of the quantum Zeno 
effect~\cite{pascazio}: the strong damping dominates the temporal 
evolution of the system and
the oscillator is frozen in its ground state $|0\rangle$, 
thus hindering the transition $|e,0\rangle \to |g,1\rangle$.
Note that $\lambda$ is the frequency of 
Rabi oscillations between levels
$|e,0\rangle$ and $|g,1\rangle$ and $\Gamma=1/\tau_d$ is the
inverse of the dissipation time scale. 
In the Zeno regime $\Gamma/\lambda\gg 1$
dissipation dominates and Rabi oscillations are suppressed.  
Our model provides a nice illustration of enhancement
of quantum channel capacities induced by the Zeno effect. 

\section{Numerical results}
In this section we discuss the results for the 
case of two channel uses. 
We numerically resolve the system dynamics related
to the Hamiltonian (\ref{model}) 
and dissipation (\ref{eq:masterdamping})
via a fourth order Runge-Kutta algorithm~\cite{diago}. 
We examine both the coherent 
information and the Holevo information and show that 
memory effects allow the channel to exceed the 
corresponding memoryless quantities. 
We mainly focus on separable two-qubit input states,
\begin{equation}
\rho_{\textsf Q_2}(0)=\rho_{\textsf Q_1}(0)^{\otimes 2}, 
\label{eq:rho2(0)}
\end{equation}
with $\rho_{\textsf Q_1}(0)$ given by Eq.~(\ref{eq:1qubit}). Entangled input states
will be briefly discussed at the end of this section.

\subsection{The coherent information}

We start our numerical investigations by computing the 
coherent information $I_c$ for the diagonal input state
\begin{equation}
\rho_{\textsf Q_2}(0)=[(1-\overline{p})|g\rangle\langle g|+
\overline{p}|e\rangle\langle e|]^{\otimes 2},
\label{eq:2qinput}
\end{equation}
where $\overline{p}$ is the value which, in the memoryless limit,
maximizes the coherent
information over single uses of the channel (see Eq. (\ref{QmemorylessCapacity})).
Since the amplitude damping channel is 
degradable~\cite{giovannettifazio}, the value obtained
from such maximization gives the quantum capacity
in the memoryless limit. 
\begin{figure}[!ht]
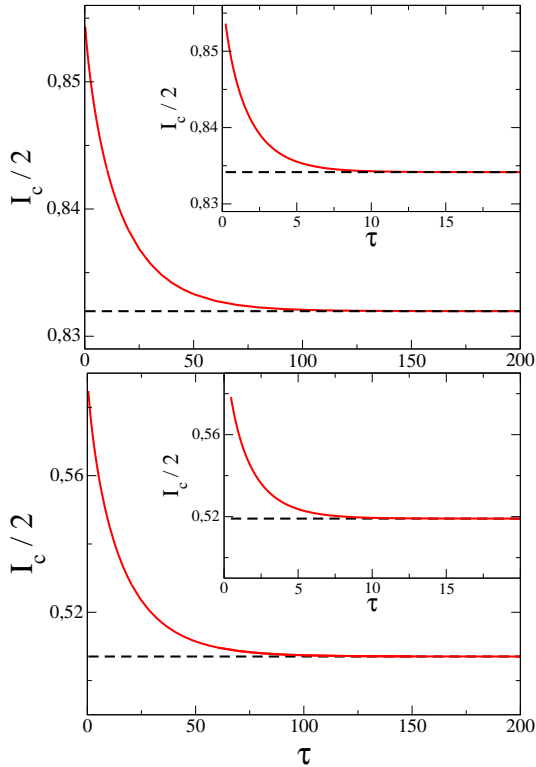

\begin{center}
\epsfig{file=fig1a.eps,width=7cm}
\epsfig{file=fig1b.eps,width=7cm}
\end{center}
\caption{Coherent information for two channel uses, at coupling strength
$\lambda=1$, $\tau_p=0.225$ (above), 
$\tau_p=0.464$ (below), 
$\Gamma=0.05$ (main plots), $\Gamma=0.5$ (insets),
$\eta=0.95$, $\overline{p}=0.4751$, (above, main plot),
$\eta=0.95$, $\overline{p}=0.4754$ (above, inset),
$\eta=0.80$, $\overline{p}=0.4490$ (below, main plot),
$\eta=0.81$, $\overline{p}=0.4497$ (below, inset).
In each plot the dashed line gives the memoryless 
quantum capacity $Q$.}
\label{fig:coherentinformation}
\end{figure}
\begin{figure}[!]
\begin{center}
\epsfig{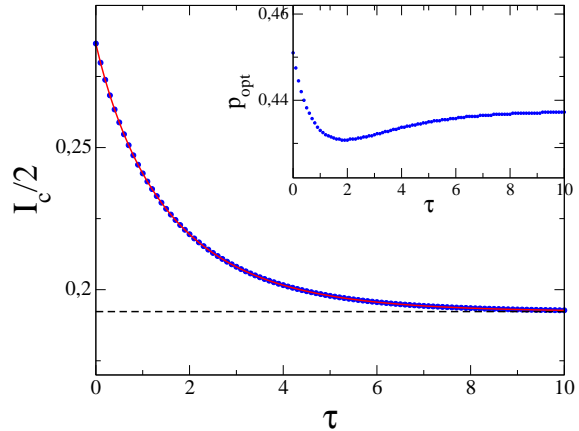}
\end{center}
\caption{Coherent information at the optimal input
state parameter $p=p_{\rm opt}$ (main plot) and 
$p_{\rm opt}$ (inset) as a function of the time 
separation $\tau$, for $\lambda=1$, $\tau_p=0.685$, $\eta=0.62$, $\Gamma=0.5$ (blue points). 
The thin red line gives the coherent information for the
value of $\overline{p}$ maximizing the memoryless limit.
The dashed line corresponds to 
the memoryless 
quantum capacity $Q$.
}
\label{fig:poptci}
\end{figure}
In Fig.~\ref{fig:coherentinformation}, we compute the 
coherent information, normalized over the number $N=2$
of channel uses, as a function of the time interval
$\tau$ between the two qubits entering the channel.
We consider weak (main plots) and strong (insets) dissipation
strength $\Gamma=1/\tau_d$ and two 
different values of the transit time $\tau_p$, leading
to higher- (top) or lower-quality (bottom) channels.
The memoryless limit is recovered in the limit $\tau\to\infty$
(memory parameter $\mu=\tau_d/(\tau+\tau_d)\to 0$), and the 
memory parameter is a decreasing function of $\tau$. 
As discussed in Sec.~\ref{sec:model},
we limit ourselves to the regime $\tau\ge \tau_p$, so that 
the maximum value of the memory parameter is 
$\tau_d/(\tau_d+\tau_p)$. 
In all plots of Fig.~\ref{fig:coherentinformation} it turns
out that the coherent information (curves) is a decreasing 
function of $\tau$, that is, a growing function of the 
memory parameter $\mu$. 
We can note that the dependence of the coherent 
information on $\tau$ (and consequently on the memory parameter $\mu$) 
is of the same kind despite the parameters $\tau_p$ and $\Gamma$ take 
very different values. 
In the same plots we show the memoryless 
quantum capacity (dashed lines).
From these plots we can conclude that 
memory effects significantly enhance the coherent 
information.
\\
The numerical optimization over the separable 
input states (\ref{eq:rho2(0)}) is achieved 
when $r=0$ (we have checked it for several values of
$\tau_p$ and $\Gamma$) and $p=p_{\rm opt}$.
As an example we show in Fig.\ref{fig:poptci}
the dependence of $p_{\rm opt}$ and 
$I_c(p_{\rm opt})$ on the time separation
$\tau$, for dissipation rate $\Gamma=0.5$.
{As one can see from Fig.\ref{fig:poptci},
even though $p_{\rm opt}$ varies as a function of $\tau$,
the improvement of the coherent information with respect to the
value of $\overline{p}$ maximizing the memoryless limit is not appreciable.}

\subsection{The Holevo information}
We numerically compute the Holevo information 
(\ref{holevo}) for separable input states:
\begin{equation}
\left\{
\begin{array}{l}
\sigma_0=|\psi_0\rangle\langle\psi_0|
\otimes |\psi_0\rangle\langle\psi_0|,
\\
\sigma_1=|\psi_0\rangle\langle\psi_0|
\otimes |\psi_1\rangle\langle\psi_1|,
\\
\sigma_2=|\psi_1\rangle\langle\psi_1|
\otimes |\psi_0\rangle\langle\psi_0|,
\\
\sigma_3=|\psi_1\rangle\langle\psi_1|
\otimes |\psi_1\rangle\langle\psi_1|,
\end{array}
\right.
\label{eq:holevosep}
\end{equation}
with 
\begin{equation}
\left\{
\begin{array}{l}
|\psi_0\rangle=\sqrt{1-\tilde{p}} |g\rangle + \sqrt{\tilde{p}} |e\rangle,
\\
|\psi_1\rangle=\sqrt{1-\tilde{p}} |g\rangle - \sqrt{\tilde{p}} |e\rangle,
\end{array}
\right.
\label{eq:psi1psi2}
\end{equation}
and equal probabilities, $\xi_k=\frac{1}{4}$ ($k=0,...,3$).
In the memoryless limit the ensemble $\{\xi_k,\sigma_k\}$ 
provides the optimal encoding strategy for $C_1$~\cite{giovannettifazio},
assuming that $\tilde{p}$ is chosen according to 
Eq.~(\ref{eq:C1memoryless}).

\begin{figure}
\begin{center}
\epsfig{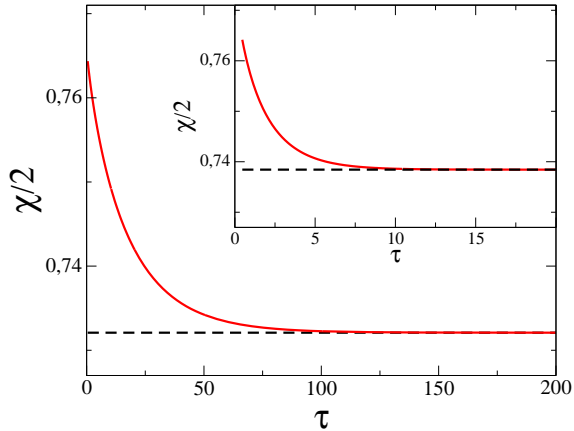}
\end{center}
\caption{Holevo information for two channel uses, at 
$\lambda=1$, $\tau_p=0.464$, 
$\Gamma=0.05$, $\eta=0.80$, $\tilde{p}=0.4329$ (main plot), 
$\Gamma=0.5$, $\eta=0.81$, $\tilde{p}=0.4338$ (inset). 
In each plot the dashed line gives the memoryless 
classical capacity $C_1$.}
\label{fig:holevoinformation}
\end{figure}

\begin{figure}
\begin{center}
\epsfig{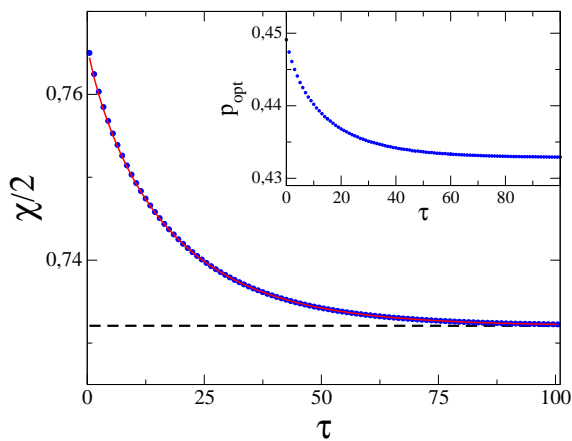}
\end{center}
\caption{Holevo information at the optimal input
state parameter $p=p_{\rm opt}$ (main plot) and
$p_{\rm opt}$ (inset) as a function of the time
separation $\tau$, for the parameter values 
of the main plot of Fig.~\ref{fig:holevoinformation}.
The thin red line is the Holevo information
for the value of $\tilde{p}$ maximizing $C_1$ in the 
memoryless limit.
The dashed line gives the memoryless
classical capacity $C_1$. 
}
\label{fig:poptholevo}
\end{figure}

In Fig.~\ref{fig:holevoinformation} we show that the behavior of the
Holevo information $\chi$ is analogous to what obtained above for the coherent
information: $\chi$ is a decreasing function of $\tau$, that is, 
a growing function function of the degree of memory.
We have then performed the optimization over the input state 
parameter $p$. The obtained $p_{\rm opt}$ and 
$\chi(p_{\rm opt})$ are shown in Fig.~\ref{fig:poptholevo}
as a function of $\tau$. 
{As for the coherent information, the Holevo quantity does not
vary appreciably by substituting the value of $p$ optimizing
the memoryless $C_1$ with $p_{opt}$.}

As it cannot generally be excluded that $C_2>C_1$ 
(or, in the case with memory, that $Q_2>Q_1$), 
we should optimize over generic two-qubit 
encodings (or input states for the quantum capacity). 
Such optimization appears difficult. In what follows, 
we limit our investigations to a simple ensemble 
whose states move, when a single parameter is varied,
from the separable ensemble 
(\ref{eq:holevosep}) to an ensemble containing 
entangled states.
We consider equal probabilities, 
$\xi_k=\frac{1}{4}$ ($k=0,...,3$) 
and states $\sigma_k=|\pi_k\rangle\langle\pi_k|$,
where
\begin{equation}
\left\{
\begin{array}{l}
|\pi_0\rangle = D_0 
(\cos\theta |\psi_0\rangle|\psi_1\rangle+
\sin\theta |\psi_1\rangle|\psi_0\rangle),
\\
|\pi_1\rangle = D_1 
(\sin\theta |\psi_0\rangle|\psi_1\rangle-
\cos\theta |\psi_1\rangle|\psi_0\rangle),
\\
|\pi_2\rangle = D_2 
(\cos\theta |\psi_0\rangle|\psi_0\rangle+
\sin\theta |\psi_1\rangle|\psi_1\rangle),
\\
|\pi_3\rangle = D_3 
(\sin\theta |\psi_0\rangle|\psi_0\rangle-
\cos\theta |\psi_1\rangle|\psi_1\rangle),
\end{array}
\right.
\label{eq:holevoent}
\end{equation}
with the states $|\psi_0\rangle$, $|\psi_1\rangle$ given 
by Eq.~(\ref{eq:psi1psi2}) and $D_k$ normalization constants.  
The separable ensemble 
(\ref{eq:holevosep}) is recovered when
$\theta= l \frac{\pi}{2} $, with $l$ integer. 
Note that in the special case $p=\frac{1}{2}$, 
$\theta = \frac{\pi}{4} + m \frac{\pi}{2}$ ($m$ integer)
the states $|\pi_i\rangle$ are maximally entangled Bell states. 

\begin{figure}
\begin{center}
\epsfig{file=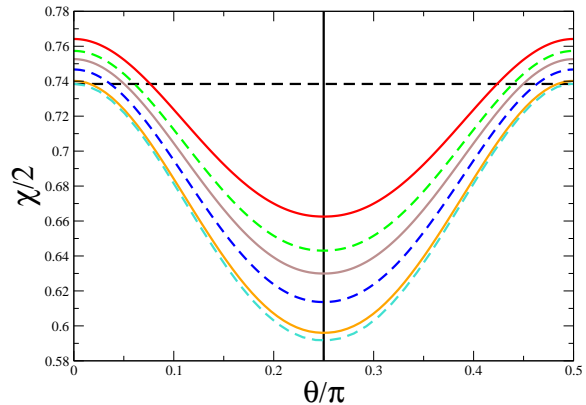,width=8.6cm}
\end{center}
\caption{Holevo information as a function of the 
parameter $\theta$ determining the ensemble 
(\ref{eq:holevoent}), for $\lambda=1$, $\tau_p=0.464$,
$\Gamma=0.5$, $\eta=0.81$, $\tilde{p}=0.4339$
and, from top to bottom, 
$\tau-\tau_p=0,0.5,1,2,5,10$.
The straight dashed line gives the memoryless capacity $C_1$.}
\label{fig:holevoent}
\end{figure}

The Holevo information as a function of $\theta$ is shown
in Fig.~\ref{fig:holevoent}, for several values of the 
separation time $\tau$. For all considered instances, we always
find that $\chi(\theta)$ is optimized by separable 
input states. Of course we cannot exclude the possibility that
different ensembles of entangled states would overcome 
the capacity $C_1$ of separable inputs.

\subsection{Memory Effects}
We discuss the mechanism which is at the heart 
of the numerically observed memory-induced enhancement of the
coherent information and the Holevo quantity.
In our memory channel model, memory not only changes the channel action at each use,
but also establishes correlations among the channel outputs. 
In this section we shall see that 
the mechanism which permits to overcome the memoryless setting is based 
on the correlations created between the two systems involved in the communication.
In particular, we will demonstrate the key role played by the entanglement
between the local environment and the information carriers.
Let us first look at quantum information transmission.
In the previous section we observed that (see Fig.~\ref{fig:coherentinformation}): 
\begin{eqnarray}
\frac{1}{2} I_c \ge  I_{c,{\rm memoryless}}.
\label{eq:memoryeffects-2}
\end{eqnarray} 
This relation approaches equality for long times $\tau$ (with respect to  $\tau_d$),
i.e. for low values of the memory parameter $\mu$.
On the other hand, the difference $\frac{1}{2} I_c -  I_{c,{\rm memoryless}}$
is larger for shorter times $\tau$,
i.e. for higher values of $\mu$.
Numerically, it turns out that the increasing of $I_c$ is mainly due to the
decreasing of the entropy exchange, since the output entropy depends very
slightly on memory.

\begin{figure}[h!]
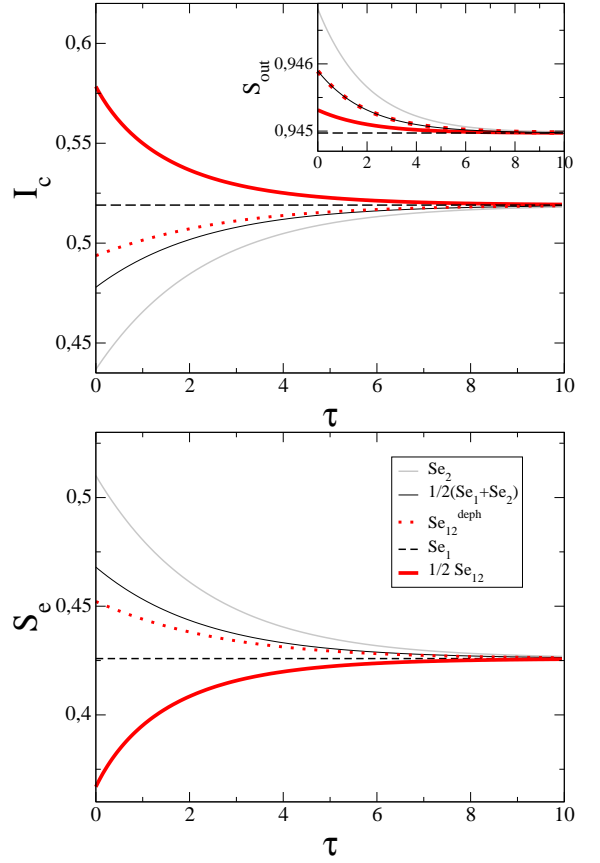

\begin{center}
\epsfig{file=fig6a.eps,width=7.5cm}
\epsfig{file=fig6b.eps,width=7.5cm}
\end{center}
\caption{Above: Coherent information (inset: output entropy) for two 
channel uses as function of the time $\tau$, at coupling strength 
 $\lambda=1$, $\tau_p=0.464$, $\Gamma=0.5$; the input state is 
 given by eq. (\ref{eq:2qinput}) with $p=0.4496$. 
 Thick red curve: 
 $\frac{1}{2}\,I_c$ (in the inset, $\frac{1}{2}S_{out}$),
 dashed black line:  
 $I_{c,1}$ ($S_{out,1}$), 
 dotted red curve:
 $\frac{1}{2}I_c^{deph}$ ($\frac{1}{2}S_{out}^{deph}$),
 black thin curve:
 $\frac{1}{2}(I_{c,1}+I_{c,2})$ ($\frac{1}{2}(S_{out,1}+S_{out,2})$),
 gray curve: 
 $I_{c,2}$ ($S_{out,2}$). 
 Note that output entropies range in an interval about fifty times smaller
 than the one relative to the coherent information.
 Below: the corresponding entropy exchanges. 
}
\label{fig:memoryeffects}
\end{figure}

At first glance, it would be tempting to say that the 
local environment entangles with the first qubit and therefore
its capability to disturb the second qubit is reduced. 
However, this is not the case. 
To understand this point we carry out a numerical analysis, in which
we consider the entropic quantities related to each single use, and compare 
them with the corresponding two-uses quantities.
We refer to $I_{c,i}$ ($S_{e,i}$, $S_{\textrm{out},i}$) as 
the coherent information (entropy exchange, output entropy) related to 
$i$-th use ($i=1,2$), 
obtained tracing out the degrees of freedom relative to the other channel use. 
We observe (see Fig.~\ref{fig:memoryeffects}) 
that the coherent information for the second channel use $I_{c,2}$
is smaller than the memoryless value $I_{c,{\rm memoryless}}$:
memory worsens the performances of the second channel use. For
the first use by definition $I_{c,1}=I_{c,{\rm memoryless}}$ 
(and $S_{e,1}=S_{e,{\rm memoryless}}$). 
So one has:
\begin{eqnarray}
\frac{1}{2}(I_{c,1}+I_{c,2}) \le  I_{c,{\rm memoryless}},
\label{eq:in_coh_1}
\end{eqnarray}
and because memory only slightly affects $S_{\textrm{out}},\,S_{\textrm{out},i}$ 
(see Fig.~\ref{fig:memoryeffects}, above, inset), inequality
(\ref{eq:in_coh_1}) is mainly due to: 
\begin{eqnarray}
\frac{1}{2} (S_{e,1}+S_{e,2}) \ge  S_{e,{\rm memoryless}}.
\label{eq:in_se_1}
\end{eqnarray}
That is, \textit{if we independently consider 
the two channel outputs, then the two uses are more noisy than in the memoryless case}. 
\begin{figure}[h!]
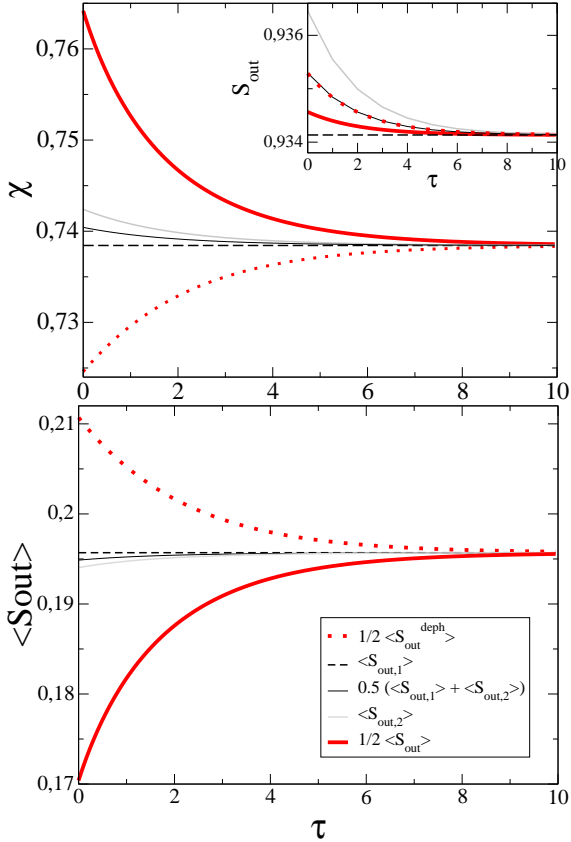

\begin{center}
\epsfig{file=fig7a.eps,width=7.5cm}
\epsfig{file=fig7b.eps,width=7.5cm}
\end{center}
\caption{Above: Holevo quantity (inset: output entropy) for two 
channel uses as function of the time $\tau$, at coupling strength 
 $\lambda=1$, $\tau_p=0.464$, $\Gamma=0.5$; the input state is 
 given by eq. (\ref{eq:holevosep}) with $p=0.4339$, $\xi_k=1/4$. 
 Thick red curve: $\frac{1}{2}\,\chi$ (in the inset, $\frac{1}{2}S_{out}$), 
 gray curve: $\chi_{2}$ ($S_{out,2}$),
 black thin curve:
 $\frac{1}{2}(\chi_{1}+\chi_{2})$ ($\frac{1}{2}(S_{out,1}+S_{out,2})$), 
 dashed black line:
 $\chi_{1}$ ($S_{out,1}$),
 dotted red curve: 
 $\frac{1}{2}\chi^{deph}$ ($\frac{1}{2}S_{out}^{deph}$). 
 Note that output entropies range in an interval about twenty times smaller
 than the one relative to the Holevo information.
 Below: the corresponding average output entropies. 
}
\label{fig:memoryeffects-classicalCap}
\end{figure}

By studying the entropy exchange behavior, we can point out
in which way memory improves the channel aptitude to transmit 
quantum information. From Fig.~\ref{fig:memoryeffects} we have that:
\begin{equation}
S_e \, \le\,S_{e,1}+S_{e,2}.
\end{equation}
It follows that correlations between systems 
${\textsf R}_1'{\textsf Q}_1'$ and ${\textsf R}_2'{\textsf Q}_2'$ satisfy:
\begin{eqnarray}
&& S({\textsf R}_1'{\textsf Q}_1':{\textsf R}_2'{\textsf Q}_2')=\nonumber\\
&&\hspace{0.5cm} S({\textsf R}_1'{\textsf Q}_1')+S({\textsf R}_2'{\textsf Q}_2')-
S({\textsf R}_1'{\textsf R}_2'{\textsf Q}_1'{\textsf Q}_2')=\nonumber\\
&&\hspace{0.5cm}S_{e,1}\,+\,S_{e,2}-S_e\,\ge\,0.
\label{eq:correlations}
\end{eqnarray}  
{(the last equality simply follows from the definitions 
$S_{e,i}=S({\textsf R}_i'{\textsf Q}_i')$ and
$S_e=S({\textsf R}_1'{\textsf R}_2'{\textsf Q}_1'{\textsf Q}_2')$).}
These correlations vanish for the memoryless setting, since by construction
systems ${\textsf R}_1{\textsf Q}_1$ and ${\textsf R}_2{\textsf Q}_2$ are initially
uncorrelated. Therefore memory effects establish correlations between
${\textsf R}_1'{\textsf Q}_1'$ and ${\textsf R}_2'{\textsf Q}_2'$; moreover
such correlations grow with the degree of memory $\mu$.
We would like to stress the key role of entanglement between the cavity and 
the qubits in establishing these correlations.
For this purpose, we consider the action of an additional source
of noise completely dephasing the local environment before the second channel
use~\cite{memorycavity}.
In this case, any entanglement between the cavity and the first qubit is destroyed.
The coherent 
information per channel use $\frac{1}{2}I_c^{\textrm{deph}}$ decreases 
(see Fig.~\ref{fig:memoryeffects}, above)
and the inequality (\ref{eq:memoryeffects-2}) 
is no longer fulfilled: the advantages produced by memory are lost.
Returning to the entropy exchange behavior we find that $S_e^\textrm{deph}$ approaches
$S_{e,1}+S_{e,2}$ so that correlations (\ref{eq:correlations}) are considerably reduced.
We can conclude that the creation of correlations between 
${\textsf R}_1'{\textsf Q}_1'$ and ${\textsf R}_2'{\textsf Q}_2'$,
mediated by qubit entanglement with the common local environment, is the essential 
mechanism allowing us to overcome the memoryless quantum capacity.

Let us look at the classical information transmission.
We recall
the Holevo information expression for a given source 
$\rho_{\textsf Q}=\sum_k \xi_k \sigma_k$:
\begin{equation} 
\chi= S(\sum_k \xi_k\sigma_k') -\sum_k \xi_k S(\sigma_k')=S_{\rm out}-\langle S_{\rm out} \rangle,
\end{equation}
where $\sigma_k'={\cal E}_2(\sigma_k)$ and 
$\langle S_{\rm out} \rangle=\sum_k \xi_k S(\sigma_k')$
is the output entropy averaged over all codewords.
We refer to $\chi_{i}$ ($\langle S_{{\rm out},i} \rangle$, $S_{{\rm out},i}$) as 
the Holevo quantity (average output entropy, output entropy) related 
to the $i$-th use. 
Differently from what we have observed for the coherent information, 
the Holevo quantity related to the second use is not worsened by memory.
But the two uses increasing
of Holevo information cannot be ascribed to the $\chi_2$ behavior. Indeed from
Fig.~\ref{fig:memoryeffects-classicalCap} (above), it appears that: 
\begin{equation}
\chi\ge\chi_1\,+\,\chi_2.
\label{eq:memory-chi-ineq}
\end{equation}
As we can see from the same figure (inset) the range of variation of the
output entropies $S_{{\rm out}}$ and $S_{{\rm out},i}$, is much smaller than the
one relative to the Holevo quantity, so that the inequality (\ref{eq:memory-chi-ineq})
is mainly due to
\begin{equation}
  \langle S_{{\rm out}} \rangle \le \langle S_{{\rm out},1} \rangle\,+\,\langle S_{{\rm out},2} \rangle.
\label{eq:memory-AvSout-ineq}
\end{equation}
We stress that the average output entropy play 
a role analogous to the entropy exchange in the quantum information transmission
scenario: it quantifies the amount of noise introduced by the channel.
We can conclude that
also from the point of view of classical information transmission, 
\textit{the channel introduces more noise
if we independently consider the two channel outputs.}
Inequality (\ref{eq:memory-AvSout-ineq}) means that, if we send 
a codeword $\sigma_i=\sigma_{i_1}\otimes \sigma_{i_2}$ 
down the channel, then the channel
establishes correlations between the output states of the two qubits.
Also in this case, these correlations are mediated by the local environment:
if we completely dephase the cavity after it interacts with the first qubit,
we lose any advantage, as one can see from Fig.~\ref{fig:memoryeffects-classicalCap}.

\section{Discussion and Conclusions}
We would like to comment on the significance of our 
results obtained with two channel uses. First of all, one
could send a couple of qubits through the channel and reset the 
oscillator to its ground state after the transit of each couple. 
In this case our results are exact and
show that for this new channel the quantum and 
classical capacities would be larger than in the memoryless
case, where the reset is operated after each channel use.
Therefore, our results can be indicative of 
a trend, that further enhancements of the coherent and 
Holevo information per channel use may be obtained by considering
longer qubit trains. 

We can also view the question from a slightly different point of view.  
We start by considering the correlations between the two-qubit system 
${\textsf Q}={\textsf Q}_1{\textsf Q}_2$ and the reference system 
${\textsf R}={\textsf R}_1{\textsf R}_2$ introduced to
purify the initial input state (indexes $1$ and $2$ refer to the first 
and the second channel use).
Such correlations are measured by the 
\emph{quantum mutual information}~\cite{groisman}
\begin{equation}
S({\textsf R}:{\textsf Q})=S({\textsf R}) + S({\textsf Q})
-S({\textsf R}{\textsf Q}),
\end{equation}
where $S$ is the von Neumann entropy and 
$S({\textsf R}{\textsf Q})$ the joint entropy for the composite system
${\textsf R}{\textsf Q}$. 
This situation describes the case in which Alice is attempting to transmit 
two halves (${\textsf Q}$) 
of two pairs of Bell states to Bob,
while she retains the other halves (${\textsf R}$). 
After the two channel uses correlations between
${\textsf R}$ and ${\textsf Q}$ read as follows:
\begin{eqnarray}
S({\textsf R}':{\textsf Q}')=S({\textsf R}') + S({\textsf Q}')
-S({\textsf R}'{\textsf Q}')
\nonumber
\\
=S({\textsf R})+S_{\rm out}-S_e= 
S({\textsf R})+I_c,
\end{eqnarray}
where the prime symbols refer to the output states, 
$S_{\rm out}=S({\textsf Q}')$ is the output entropy 
and $S_e=S({\textsf R}'{\textsf Q}')$ the entropy exchange. 
Since $S({\textsf R}')=S({\textsf R})$ is independent
of the degree of memory $\mu$ (i.e., independent of the separation 
time $\tau$), we find that the fact that memory increases  
coherent information $I_c$ is deeply related to the fact that 
memory helps to preserve correlations between 
${\textsf R}$ and ${\textsf Q}$.

With regard to the Holevo information, a similar argument 
can be developed. Indeed, we can model the mixed classical-quantum
ensemble $\{\xi_k,\sigma_k\}$ by means of an enlarged Hilbert space 
representation~\cite{devetak}:
\begin{equation}
\rho_{{\textsf A}{\textsf Q}}=
\sum_k \xi_k |k\rangle\langle k| \otimes \sigma_k,
\end{equation}
where $\{|k\rangle\}$ is and orthonormal basis for the
Hilbert space of the auxiliary system ${\textsf A}$.
The correlations between ${\textsf A}$ 
and ${\textsf Q}$ read
\begin{eqnarray}
S({\textsf A}:{\textsf Q})&=& 
S({\textsf A})+S({\textsf Q})-S({\textsf A}{\textsf Q})= \nonumber\\
&&\hspace{-0.0cm} S({\textsf Q})-S({\textsf Q}|{\textsf A})= \nonumber\\
&&\hspace{-0.0cm} S(\sum_k \xi_k\sigma_k) -\sum_k \xi_k S(\sigma_k)
\label{holevoenlarged}
\end{eqnarray}
where $S({\textsf Q}|{\textsf A})=S({\textsf A}{\textsf Q})
-S({\textsf A})$ is a conditional von Neumann entropy.
Now if we choose an ensemble of pure states $\sigma_k$, the
correlations between the systems ${\textsf A}$ and ${\textsf Q}$ are maximum,
and equal to $S({\textsf A})=S(\sum_k \xi_k\sigma_k)$, since $S(\sigma_k)=0$.
But when we transmit the quantum system ${\textsf Q}$ down the channel, it adds
noise such that $\sigma_k \rightarrow \sigma_k'={\cal E}_2(\sigma_k)$,
and in general  
$S(\sigma_k')\neq 0$. Thus 
correlations between ${\textsf A}$ and ${\textsf Q}$ decrease to:
\begin{eqnarray}
 \hspace{-1.2cm} S({\textsf A}:{\textsf Q'}) &=& S({\textsf Q'})-S({\textsf Q'}|{\textsf A})= \nonumber\\
 && \hspace{-1.4cm}S(\sum_k \xi_k\sigma_k') -\sum_k \xi_k S(\sigma_k')= \chi(\{\xi_k,\sigma_k\})
\label{holevoenlarged-2}
\end{eqnarray}
Then the Holevo information is just the correlation between the classical variable
we want to transmit and the quantum system we have used to transmit it.
Since memory enhances the Holevo quantity with respect to the memoryless
setting, we conclude that 
memory helps to preserve these correlations.

In both scenarios - quantum and classical information transmission -
memory helps in preserving the correlations between the systems 
involved in the transmission. We stress 
again the key role of entanglement between information carriers and the 
channel.
This effect could be in principle useful for quantum error-correcting codes
(QECC).
Indeed, understanding in which way memory acts during information transmission,
may suggest suitable strategies to design new QECC with performances better 
than those of standard memoryless QECC. 
{In this perspective, further work will be necessary to see if,
in presence of memory,
the coherent information and Holevo quantity per channel 
use increase with the number of channel uses, as one would expect from the data  
shown in the Figs.~\ref{fig:coherentinformation}$\,-\,$\ref{fig:holevoinformation}, 
as it happens for the memory dephasing channel~\cite{njp}}.

\section*{Acknowledgements}
We acknowledge fruitful discussions with Shash Virmani
and an anonymous referee for useful remarks.
AD acknowledges L. Bordi, T. D'Arrigo and M. Fici for invaluable help.
This work was partially supported by the EU through grant
no. PITN-GA-2009-234970, and by the Joint Italian-Japanese
Laboratory on ``Quantum Technologies: Information, Communication and
Computation" of the Italian Ministry of Foreign Affairs.

\appendix

\section{Quantum capacity of the channel $\overline{\cal E}_{N+L}$}
\label{app:doubleBlockStrategy}
We discuss the condition under which the quantum capacity of the channel
$\bar{\mathcal{E}}_{N+L}$ is the same as for the $\mathcal{E}_{N+L}$. 
We consider $n=N+L$ channel use; the related coherent information is:
\begin{eqnarray}
&&\hspace{-1cm} I_c(\mathcal{E}_n,\rho^{(n)})\,=\nonumber\\
&&\hspace{-0.5cm} S[\mathcal{E}_n(\rho^{(n)})]-
S[({\cal I}^{(n)} \otimes \mathcal{E}_n)(|\Psi^{(n)}\rangle\langle\Psi^{(n)}|)].
\label{appx-dbs-1}
\end{eqnarray}
where $|\Psi^{(n)}\rangle$ is any purification of $\rho^{(n)}$.
Now we call ${\textsf Q}^{(N)}$  and ${\textsf Q}^{(L)}$ the systems composed by 
the first $N$ and the last $L$ qubits, respectively. We refer to
${\textsf R}$ as the purifying system.
We observe that:
\begin{eqnarray}
  && \hspace{-1cm} \mathcal{E}_n(\rho^{(n)})\,=
  \,\big(({\cal I}^{(N)}\otimes {\cal N}_L) \circ (\mathcal{E}_N\otimes {\cal I}^{(L)})\big) (\rho^{(N+L)}); 
\label{appx-dbs-2}
\end{eqnarray}
indeed we can image that the overall channel processing on a $n$-input state, can be 
decomposed in two steps. First, we operate only on the first $N$ use, leaving 
unaffected the successive $L$ uses (${\cal I}_L$ is the identity operator): 
this corresponds to act on the first $N$-uses by
the quantum map ${\cal E}_N$.
Second, we perform the quantum operation ${\cal N}_L$ on the remaining $L$-use,
leaving unaffected the first ones, in a such way the overall action on the
$n$-input state is still described by ${\cal E}_n$.
By the \textit{data processing inequality}~\cite{barnum} we have:
\begin{eqnarray}
&&I_c(\mathcal{E}_n,\rho^{(n)})\,\le I_c(\mathcal{E}_N\otimes {\cal I}^{(L)},\rho^{(n)}).
\label{appx-dbs-3}
\end{eqnarray}
By definition of coherent information~\cite{schumachernielsen},
\begin{eqnarray}
&& \hspace{-0.7cm}I_c(\mathcal{E}_N\otimes {\cal I}^{(L)},\rho^{(n)})\,=\nonumber\\
&& \hspace{0.0cm}S\big({\cal E}_N\otimes {\cal I}^{(L)}(\rho^{(n)})\big)\,-\nonumber\\
&& \hspace{0.2cm}S\big({\cal I}^{(n)}\otimes ({\cal E}_N\otimes {\cal I}^{(L)})|\Psi^{(n)}\rangle\langle\Psi^{(n)}|\big),
\label{appx-dbs-4}
\end{eqnarray}
we observe that the last term in the (\ref{appx-dbs-4}) is just the entropy
exchange~\cite{schumacher} related to the first $N$ channel uses map ${\cal E}_N$.
Indeed $|\Psi^{(n)}\rangle$, being a purification of $\rho^{(n)}$, it also is 
a purification of $\rho^{(N)}=\mathrm{Tr}_{{\textsf Q}^{(L)}}\{\rho^{(n)}\}$,
and the map ${\cal I}^{(n)}\otimes ({\cal E}_N\otimes {\cal I}^{(L)})$
acts on ${\textsf Q}^{(N)}$ by ${\cal E}_N$, while does not change the purifying systems 
${\textsf R}$ and ${\textsf Q}^{(L)}$.
Moreover, by the subadditivity of the von Neumann entropy we can bound from above
the first term at the second member of (\ref{appx-dbs-4}): 
\begin{eqnarray}
  && \hspace{-1cm} S\big({\cal E}_N\otimes {\cal I}^{(L)}(\rho^{(n)})\big) \le \nonumber\\
  && \hspace{-0.5cm} S\big(\mathrm{Tr}_{{\textsf Q}^{(L)}}\{{\cal E}_N\otimes {\cal I}^{(L)}(\rho^{(n)})\}\big)\,+\nonumber\\
  && \hspace{-0.0cm} S\big(\mathrm{Tr}_{{\textsf Q}^{(N)}}\{{\cal E}_N\otimes {\cal I}^{(L)}(\rho^{(n)})\}\big)\,=\, \nonumber\\
  && \hspace{-0.5cm} S\big({\mathcal{E}}_{N}(\rho^{(N)})\big)\,+\,
     S\big(\mathrm{Tr}_{{\textsf Q}^{(N)}}\{{\cal E}_N\otimes {\cal I}^{(L)}(\rho^{(n)})\}\big)\,\le \nonumber\\
  && \hspace{-0.5cm} S\big({\mathcal{E}}_{N}(\rho^{(N)})\big)\,+\, L\log_2 \dim({\cal H}_1),
\label{appx-dbs-5}
\end{eqnarray}
where  
we used \textit{causality}~\cite{werner} in taking the trace with respect to ${\textsf Q}^{(L)}$: 
outputs related to the first $N$ uses do not depend on inputs 
related to the subsequent $L$ uses. 
So, returning to the coherent information (\ref{appx-dbs-4}), we can write:
\begin{eqnarray}
&& \hspace{-0.7cm}I_c(\mathcal{E}_N\otimes {\cal I}^{(L)},\rho^{(n)})\,\le \nonumber\\
&&  S\big({\mathcal{E}}_{N}(\rho^{(N)})\big)\,+\, L\log_2 \dim({\cal H}_1)\,-\nonumber\\
&& \hspace{0.2cm}S\big({\cal I}^{(n)}\otimes ({\cal E}_N\otimes {\cal I}^{(L)})|\Psi^{(n)}\rangle\langle\Psi^{(n)}|\big)\,=\nonumber\\
&& \hspace{0.2cm} I_c(\mathcal{E}_N,\rho^{(N)})\,+\, L\log_2 \dim({\cal H}_1)
\label{appx-dbs-6}
\end{eqnarray}
Finally by the (\ref{appx-dbs-3}) and (\ref{appx-dbs-6}) we have that:
\begin{eqnarray}
&& I_c(\rho^{(N+L)},{\cal E}_{N+L}) \le \nonumber \\
&&\hspace{1cm} I_c(\bar{\mathcal{E}}_{N+L},\rho^{(N+L)}) \, + \, L\log_2 \dim({\cal H}_1)
\label{appx-dbs-7}
\end{eqnarray}
since, by the definition of double blocking strategy,  
$$I_c(\bar{\mathcal{E}}_{N+L},\rho^{(N+L)})=I_c(\mathcal{E}_N,\rho^{(N)}).$$
By maximizing over all possible input $\rho^{(n)}=\rho^{(N+L)}$, taking the limit
for $N \to \infty$ and assuming $\lim_{N \to \infty} L/N=0$, (\ref{appx-dbs-7}) assures that
$Q({\mathcal{E}}_{N+L}) \le Q(\bar{\mathcal{E}}_{N+L})$. On the other
hand, the capacity of the 
channel $\mathcal{E}_{N+L}$ is obviously at least as great as the one of $\bar{\mathcal{E}}_{N+L}$,
so one can conclude
that $Q({\mathcal{E}}_{N+L}) = Q(\bar{\mathcal{E}}_{N+L})$.

\section{Proof of eq. (\ref{eq:forgetful}) for the channel (\ref{model})}
\label{app:forgetfulness}
We consider $M$ blocks of $N+L$ uses of the channel (\ref{model}), in a 
double blocking strategy perspective (see section \ref{sec:memorychannel}).
We want to show that: 
\begin{eqnarray}
\Vert \bar{\mathcal{E}}_{M(N+L)}(\rho_{\textsf Q}) -
\bar{\mathcal{E}}_{N+L}^{\otimes M}(\rho_{\textsf Q}) \Vert_1
\leq h (M-1)c^{-L}.
\label{appx-B:1}
\end{eqnarray}
We suppose to deal with a larger number $M'>M$ of blocks, whose
the first $M$ are subjected to the channels operations $\bar{\mathcal{E}}_{M(N+L)}$ or
$\bar{\mathcal{E}}_{N+L}^{\otimes M}$, whereas the remaining
$M'-M$ do not undergo any operation. Note that $M'$ is just introduced for technical
purpose, in order to carry out the proof. 
Thanks to the contractivity of trace-preserving quantum 
operations~\cite{nielsen-chuang}, we can write:
\begin{eqnarray}
&&\hspace{-1.5cm} \Vert \bar{\mathcal{E}}_{M(N+L)}(\rho_{\textsf Q}) -
\bar{\mathcal{E}}_{N+L}^{\otimes M}(\rho_{\textsf Q}) \Vert_1 \leq \nonumber\\
&&\hspace{-1cm}\Vert {\cal I}_{(M'-M)(N+L)}\otimes\bar{\mathcal{E}}_{M(N+L)}(\rho_{\textsf Q}) -\nonumber\\
&&\hspace{1cm}{\cal I}_{(M'-M)(N+L)}\otimes \bar{\mathcal{E}}_{N+L}^{\otimes M}(\rho_{\textsf Q})\Vert_1.
\label{appx-B:1a}
\end{eqnarray}
By further using the contractivity, we can upper bound from above the right part of the  
(\ref{appx-B:1a}) by:
\begin{eqnarray}
&& \hspace{-0.7cm} \Vert \tilde{{\cal I}}_{(M'-M)(N+L)}\otimes\bar{\mathcal{F}}_{M(N+L)}(\rho_{\textsf Q}\otimes \rho_{\textsf O}) -\nonumber\\
&& \hspace{0.7cm} \tilde{{\cal I}}_{(M'-M)(N+L)}\otimes\bar{\mathcal{F}}_{N+L}^{\otimes M}(\rho_{\textsf Q}\otimes 
\rho_{\textsf O}) \Vert_1,
\label{appx-B:2}
\end{eqnarray}
where $\tilde{{\cal I}}$ is the identity operator acting on the composite 
system $\textsf{QO}$, $\rho_{\textsf O}=|0\rangle\langle 0|$ is the 
ground state of the oscillator and, as we have seen in section \ref{sect:forgetfulness},
the quantum operations $\bar{\mathcal{F}}_{j(N+L)}$ are such that, 
for any integer $j$, 
\begin{equation}
\bar{\mathcal{E}}_{j(N+L)}(\rho_{\textsf Q})=
{\rm Tr}_{\textsf O}[\bar{\mathcal{F}}_{j(N+L)}
(\rho_{\textsf Q}\otimes \rho_{\textsf O})].
\label{appx-B:3}
\end{equation}
The memoryless map $\bar{\mathcal{F}}_{N+L}^{\otimes M}$ corresponds to apply
$M$ times the map $\bar{\mathcal{F}}_{N+L}$ after the state of the oscillator
has been reset to its ground state. 
Due to the triangle inequality we can bound from above the (\ref{appx-B:2}) by:
\begin{eqnarray}
&&\hspace{-0.5cm}\Vert \tilde{{\cal I}}_{(M'-M)(N+L)}\otimes\bar{\mathcal{F}}_{M(N+L)}(\rho_{\textsf Q}\otimes \rho_{\textsf O}) -
   \label{appx-B:4}\\
&&\hspace{-0.2cm}      \tilde{{\cal I}}_{(M'-M)(N+L)}\otimes\bar{\mathcal{F}}_{N+L}^{(\pi_0)}\otimes\bar{\mathcal{F}}_{(M-1)(N+L)}(\rho_{\textsf Q}\otimes \rho_{\textsf O})\Vert_1+ \nonumber\\
&&\hspace{-0.5cm}\Vert \tilde{{\cal I}}_{(M'-M)(N+L)}\otimes\bar{\mathcal{F}}_{N+L}^{(\pi_0)}\otimes\bar{\mathcal{F}}_{(M-1)(N+L)}(\rho_{\textsf Q}\otimes \rho_{\textsf O})\label{appx-B:5}\\
&&\hspace{-0.2cm} -\tilde{{\cal I}}_{(M'-M)(N+L)}\otimes \bar{\mathcal{F}}_{N+L}^{\otimes M}(\rho_{\textsf Q}\otimes \rho_{\textsf O})
   \Vert_1.\nonumber
\end{eqnarray}
where we set $\bar{\mathcal{F}}_{N+L}^{(\pi_0)}\equiv \bar{\mathcal{F}}_{N+L}\circ\,\pi_0$,
being $\pi_0$ the quantum operation deleting the qubits-cavity correlations and 
resetting the cavity in its 
ground state (\ref{eq:memoryless-channel-resetting}). 
Let us consider the trace norm (\ref{appx-B:4}). The first $(M-1)(N+L)$ uses of
the two channels in (\ref{appx-B:4}) are the same, bringing the 
system $\textsf{QO}$ in the state $\rho_\textsf{QO}\prime$, so we can rewrite the 
(\ref{appx-B:4}) as:
\begin{eqnarray}
&&\hspace{-0.5cm}\Vert \tilde{{\cal I}}_{(M'-M)(N+L)}\otimes \bar{\mathcal{F}}_{N+L}(\rho_{\textsf{QO}}^\prime) -\nonumber \\
&&\hspace{+0.5cm}\tilde{{\cal I}}_{(M'-M)(N+L)}\otimes\bar{\mathcal{F}}_{N+L}[\pi_0 (\rho_{\textsf{QO}}^\prime)]\Vert_1.
\label{appx-B:6}
\end{eqnarray}
As we have seen in section \ref{sect:forgetfulness} for  
the expression in  (\ref{eq:forgetful4}), it can be shown that the trace norm 
(\ref{appx-B:6}) is smaller than: 
\begin{equation}
\Vert \rho_{\textsf{QO}}^\prime -\rho_{\textsf Q}'\otimes 
|0\rangle\langle 0| \Vert_1
< 4 \sqrt{B} e^{-L\Gamma\tau/2}.
\end{equation}
Now we turn to the trace norm (\ref{appx-B:5}). The last not trivial $N+L$ uses for the
two channels in the (\ref{appx-B:5}) are described by the same quantum operation,
so that by applying the contractivity of the trace norm with respect a quantum operation,
we can upper bound the (\ref{appx-B:5}) by:
\begin{eqnarray}
&&\hspace{-0.5cm} \Vert \tilde{{\cal I}}_{(M'-M+1)(N+L)}\otimes\bar{\mathcal{F}}_{(M-1)(N+L)}(\rho_{\textsf Q}\otimes \rho_{\textsf O})-\nonumber\\
&&\hspace{-0.1cm}\tilde{{\cal I}}_{(M'-M+1)(N+L)}\otimes \bar{\mathcal{F}}_{N+L}^{\otimes (M-1)}(\rho_{\textsf Q}\otimes \rho_{\textsf O})\Vert_1
\label{appx-B:7}
\end{eqnarray} 
Note that this expression is formally equivalent to (\ref{appx-B:2}), indeed 
it can be derived by the 
(\ref{appx-B:2}) simply by replacing $M$ with $M-1$. 
By summarizing we have bounded the (\ref{appx-B:2}) in the following manner:
\begin{eqnarray}
&& \hspace{-1cm} \Vert \tilde{{\cal I}}_{(M'-M)(N+L)}\otimes\bar{\mathcal{F}}_{M(N+L)}(\rho_{\textsf Q}\otimes \rho_{\textsf O}) -\nonumber\\
&& \hspace{-0.5cm}  \tilde{{\cal I}}_{(M'-M)(N+L)}\otimes\bar{\mathcal{F}}_{N+L}^{\otimes M}(\rho_{\textsf Q}\otimes 
\rho_{\textsf O}) \Vert_1 \,\le \nonumber\\
&&\hspace{-1cm}4 \sqrt{B} e^{-L\Gamma\tau/2}+\nonumber\\
&&\hspace{-0.5cm} \Vert \tilde{{\cal I}}_{(M'-M+1)(N+L)}\otimes\bar{\mathcal{F}}_{(M-1)(N+L)}(\rho_{\textsf Q}\otimes \rho_{\textsf O})-\nonumber\\
&&\hspace{-0.0cm}\tilde{{\cal I}}_{(M'-M+1)(N+L)}\otimes \bar{\mathcal{F}}_{N+L}^{\otimes (M-1)}(\rho_{\textsf Q}\otimes \rho_{\textsf O})\Vert_1
\label{appx-B:8}
\end{eqnarray}
By recursively applying  the (\ref{appx-B:8}), up to $M-3$ times, we obtain: 
\begin{eqnarray}
&& \hspace{-1cm} \Vert \tilde{{\cal I}}_{(M'-M)(N+L)}\otimes\bar{\mathcal{F}}_{M(N+L)}(\rho_{\textsf Q}\otimes \rho_{\textsf O}) -\nonumber\\
&& \hspace{-0.5cm}  \tilde{{\cal I}}_{(M'-M)(N+L)}\otimes\bar{\mathcal{F}}_{N+L}^{\otimes M}(\rho_{\textsf Q}\otimes 
\rho_{\textsf O}) \Vert_1 \,\le \nonumber\\
&&\hspace{-1cm}4  \sqrt{B} (M-2) e^{-L\Gamma\tau/2}+\nonumber\\
&&\hspace{-0.5cm} \Vert \tilde{{\cal I}}_{(M'-2)(N+L)}\otimes\bar{\mathcal{F}}_{2(N+L)}(\rho_{\textsf Q}\otimes \rho_{\textsf O})-\nonumber\\
&&\hspace{-0.0cm}\tilde{{\cal I}}_{(M'-2)(N+L)}\otimes \bar{\mathcal{F}}_{N+L}^{\otimes (2)}(\rho_{\textsf Q}\otimes \rho_{\textsf O})\Vert_1
\label{appx-B:9}
\end{eqnarray}
As for the (\ref{eq:forgetful3}), it can be shown that the last trace norm in 
(\ref{appx-B:9}) is smaller than $4 \sqrt{B} e^{-L\Gamma\tau/2}$, so that:
\begin{eqnarray}
&& \hspace{-1cm} \Vert \tilde{{\cal I}}_{(M'-M)(N+L)}\otimes\bar{\mathcal{F}}_{M(N+L)}(\rho_{\textsf Q}\otimes \rho_{\textsf O}) -\nonumber\\
&& \hspace{-0.5cm}  \tilde{{\cal I}}_{(M'-M)(N+L)}\otimes\bar{\mathcal{F}}_{N+L}^{\otimes M}(\rho_{\textsf Q}\otimes 
\rho_{\textsf O}) \Vert_1 \,\le \nonumber\\
&&\hspace{-1cm}4 \sqrt{B} (M-1) e^{-L\Gamma\tau/2} 
\label{appx-B:10}
\end{eqnarray}
Because, as we say above, the left part of (\ref{appx-B:1}) can be upper bounded
by (\ref{appx-B:2}), we can conclude that:
\begin{eqnarray}
&& \hspace{-1.5cm}\Vert \bar{\mathcal{E}}_{M(N+L)}(\rho_{\textsf Q}) -
\bar{\mathcal{E}}_{N+L}^{\otimes M}(\rho_{\textsf Q}) \Vert_1 \leq \nonumber\\
&& \hspace{1cm} 4 (M-1) \sqrt{B} e^{-L\Gamma\tau/2}.
\label{appx-B:11}
\end{eqnarray}
\vspace*{0.1cm}

\section{Map for a single channel use}
\label{app:master1q}
We solve the Lindblad master equation (\ref{eq:Lindblad1q}),
with the qubit prepared in the generic input state 
$\rho_1{(0)}$ of Eq.~(\ref{eq:1qubit})
and the oscillator initially in its ground state. 
During time evolution the overall system (qubit plus oscillator)
resides in a three-dimensional subspace, spanned by the states
$|0\rangle\equiv |g,0\rangle$,
$|1\rangle\equiv |g,1\rangle$,
$|2\rangle\equiv |e,0\rangle$.
We define $\rho_{ij}\equiv 
\langle i | \rho_{\textsf{QO}} | j\rangle$ ($i,j=0,1,2$)
and write the master equation in the $\{|0\rangle,|1\rangle,|2\rangle\}$ 
basis. We obtain the following differential equations:
\begin{eqnarray}
\dot{\rho}_{00}=\Gamma \rho_{11},
\\
\dot{\rho}_{01}=i\lambda \rho_{02}-\frac{\Gamma}{2}\rho_{01},
\\
\dot{\rho}_{02}=i\lambda \rho_{01},
\\
\dot{\rho}_{11}=i\lambda(\rho_{12}-\rho_{21})-\Gamma\rho_{11},
\\
\dot{\rho}_{12}=i\lambda(\rho_{11}-\rho_{22})-\frac{\Gamma}{2}\rho_{12},
\\
\dot{\rho}_{22}=-i\lambda(\rho_{12}-\rho_{21}),
\end{eqnarray}
with initial conditions $\rho_{00}(t=0)=1-p$, $\rho_{01}(0)=0$,
$\rho_{02}(0)=r$, $\rho_{11}(0)=0$, $\rho_{12}(0)=0$, $\rho_{22}(0)=p$.
These linear equations can be solved in a standard way using the Laplace 
transform~\cite{chen}. After tracing the solutions at time $\tau_p$ over the
oscillator, we obtain the CPT map
\begin{equation}
\mathcal{E}_1[\rho_1{(0)}]=
\left[
\begin{array}{cc}
 1-p[h(\Gamma)]^2 & r h(\Gamma)\\
 r^\star h(\Gamma) & p [h(\Gamma)]^2
\end{array}
\right],
\end{equation}
where 
\begin{equation}
h(\Gamma)=e^{-\Gamma \tau_p/4}
\left[\frac{\Gamma+ z}{2z} e^{z\tau_p/4}-
\frac{\Gamma-z}{2z} e^{-z\tau_p/4}\right],
\end{equation}
with $z=\sqrt{\Gamma^2-16\lambda^2}$.
This is an amplitude-damping channel 
as in Eq.~(\ref{eq:amplitudedamping}), with parameter 
$\eta(\Gamma)=[h(\Gamma)]^2$.


\begin{thebibliography}{}
\bibitem{notenorm}
Inequality (\ref{psfcondition4}) implies (\ref{psfcondition2})
since (see~\cite{nielsen-chuang}) 
\begin{equation}
|||\psi\rangle\langle \psi| - \sigma||_1
\ge 2(1-\sqrt{\langle \psi |\sigma |\psi\rangle}),
\end{equation}
where we have defined 
$\sigma=\bar{\mathcal{D}}^{(M)}\circ
\bar{\mathcal{E}}_{N+L}^{\otimes M}
\circ \bar{\mathcal{C}}^{(M)}
(|\psi\rangle\langle \psi|)$.
\bibitem{noteM0}
The quantity (\ref{psfcondition3}) is bounded from above by the sum of 
(the right side of) (\ref{psfcondition4}) and (\ref{psfcondition5}). By hypothesis 
the right side of (\ref{psfcondition4}) is less or equal than $\varepsilon/2$ when $M>M_0$.
The quantity (\ref{psfcondition5}) is bounded by $h(M-1)c^{-L}$;
because of $\lim_{M \to \infty} h(M-1)c^{-\varepsilon M}=0$, it there exists $\tilde{M}$
such that $\forall M>\tilde{M} \quad \Rightarrow \quad h(M-1)c^{-\varepsilon M}\le \varepsilon/2$.
Then $M_0'= \max\{M_0,\tilde{M}\}$.
\end{thebibliography}
\end{document}